\documentclass[review]{elsarticle}

\usepackage{lineno,hyperref,fullpage}
\usepackage{afterpage}
\usepackage{enumitem}
\usepackage{snippet}
\usepackage{siunitx}
\usepackage{booktabs}
\usepackage{longtable}
\usepackage{graphicx}
\usepackage{subfigure}
\usepackage{changes}
\usepackage[ruled,vlined]{algorithm2e}

\journal{Elsevier}

\usepackage{nomencl}
\usepackage{amsmath}
\usepackage{amsfonts}
\usepackage{multirow}
\usepackage{subcaption}

\usepackage{floatrow}
\makenomenclature
\usepackage{xpatch}
\xpatchcmd{\thenomenclature}{\section*{\nomname}
}{}{\typeout{Success}}{\typeout{Failure}}
\makenomenclature
\RequirePackage{ifthen}
\printnomenclature[0.8cm]

\usepackage{soul}
\usepackage{xcolor}
\usepackage{bm}
\usepackage{url}
\usepackage{courier}

\renewcommand\nomgroup[1]{
  \item[\itshape
  \ifstrequal{#1}{A}{Symbols}{
  \ifstrequal{#1}{B}{Roman Letters}{
  \ifstrequal{#1}{C}{Greek Letters}{
  \ifstrequal{#1}{D}{Abbreviations}{}}}}
]}

\let\oldequation\equation
\let\oldendequation\endequation
\renewenvironment{equation}
  {\linenomathNonumbers\oldequation}
  {\oldendequation\endlinenomath}

\begin{document}

\begin{frontmatter}
\title{A framework for learning symbolic turbulence models from indirect observation data via neural networks and feature importance analysis}

  \author[cas,ucas]{Chutian Wu}
  \author[cas,ucas]{Xin-Lei Zhang\corref{mycor}}
  \ead{zhangxinlei@imech.ac.cn}
  \author[cas,ucas]{Duo Xu}
  \author[cas,ucas]{Guowei He}
  \cortext[mycor]{Corresponding author}

  \address[cas]{State Key Laboratory of Nonlinear Mechanics, Institute of Mechanics, Chinese Academy of Sciences, Beijing, 100190, China}
  \address[ucas]{School of Engineering Sciences, University of Chinese Academy of Sciences, Beijing 100049, China}

  \begin{abstract}
    Learning symbolic turbulence models from indirect observation data is of significant interest as it not only improves the accuracy of posterior prediction but also provides explicit model formulations with good interpretability.
    However, it typically resorts to gradient-free evolutionary algorithms, which can be relatively inefficient compared to gradient-based approaches, particularly when the Reynolds-averaged Navier-Stokes (RANS) simulations are involved in the training process.
    In view of this difficulty, we propose a framework that uses neural networks and the associated feature importance analysis to improve the efficiency of symbolic turbulence modeling.
    In doing so, the gradient-based method can be used to efficiently learn neural network-based representations of Reynolds stress from indirect data, which is further transformed into simplified mathematical expressions with symbolic regression.
    Moreover, feature importance analysis is introduced to accelerate the convergence of symbolic regression by excluding insignificant input features.
    The proposed training strategy is tested in the flow in a square duct, where it correctly learns underlying analytic models from indirect velocity data.
    Further, the method is applied in the flow over the periodic hills, demonstrating that the feature importance analysis can significantly improve the training efficiency and learn symbolic turbulence models with satisfactory generalizability.
  \end{abstract}

  \begin{keyword}
    turbulence model \sep ensemble Kalman method \sep symbolic regression \sep feature importance analysis \sep neural networks
  \end{keyword}
\end{frontmatter}


\section{Introduction}
Data-driven approaches have emerged as an alternative tool for augmenting turbulence models, thereby improving the predictive accuracy of the Reynolds-averaged Navier-Stokes (RANS) simulations. 
 Different observation data, such as the indirect data of velocity~\cite{holland2019, STROFER2021TAML} and the direct data associated with Reynolds stress~\cite{ling2016reynolds, Wu2017PhysRevFluids}, have been used to discover optimal functional mapping between mean flows to the Reynolds stress.
 Indirect observation data is relatively straightforward to obtain compared to direct data, particularly for flows with high Reynolds numbers.
 Moreover, the training strategies based on indirect data can ensure good predictive accuracy of mean flow fields as the RANS equation is involved during the training process~\cite{holland2019, zhang2022, Liu2023}.
 Therefore, it becomes of significant interest to learn accurate turbulence models from indirect observation data.
 
 Indirect data have been used to learn turbulence models represented in various forms, including neural networks~\cite{ling2016reynolds, singh2017machine} and symbolic expression~\cite{WEATHERITT201622}. 
Neural networks are essentially composite nonlinear functions, which can have great expressive power based on the well-known universal approximation theory~\cite{pinkus1999approximation,zhang2024multi}.
In contrast, the symbolic expression represents a mathematical formula with various alphanumeric characters and operation symbols arranged in specific rules~\cite{makke2024interpretable}.
 Both model representations have been demonstrated for improving the accuracy of the RANS simulations based on indirect observation data~\cite{holland2019,ZHAO2020,STROFER2021TAML,zhang2022}.
In the following, we discuss briefly the two types of model forms in the model interpretability and training efficiency.

The neural network-based method can efficiently learn complicated model forms of Reynolds stress in a black box.
 Specifically, neural networks have great expressive power to characterize very complex mathematical functions with multiple layers. 
Moreover, the gradient-based method can be applied to efficiently optimize neural network weights based on indirect observation data.
 However, the learned neural networks are black-box models due to multiple compositions of nonlinear functions, leading to the difficulty in model interpretability.
 This intrinsic attribute of the neural-network-based turbulence model results in a relatively low rank in the model readiness rating (MRR) system~\cite{NASA}. 
Several works have been conducted to enhance the interpretability of the neural network-based turbulence model, mainly by analyzing the feature importance. 
For instance, the SHAP (SHapley Additive exPlanations)~\cite{HE2022,WU2023} and the permutation feature importance (PFI) methods~\cite{MANDLER2023} have been used to indicate the contribution of each input feature to model predictions.
However, these model-agnostic and post-hoc methods only offer partial interpretability for the neural network-based turbulence models, i.e., the relative importance of input features.
As such, the symbolic expression gains increasing interest due to its good interpretability and ease of further model development.

The symbolic expression-based method can provide an analytical formulation of Reynolds stress in a white box.
It aims to yield simple mathematical expressions of the Reynolds stress that improve the RANS prediction and have good physical interpretability.
Various symbolic regression methods have been developed for learning turbulence models based on indirect observation data.
For instance, the gene expression programming (GEP) approach~\cite{WEATHERITT201622} has been utilized to learn symbolic expressions of the Reynolds stress from various indirect data~\cite{ZHAO2020, Fang2023, LAV2023109140}.
It is achieved by solving the RANS equations coupled with different candidate models to evaluate corresponding cost functions. 
Alternatively, the symbolic regression can be coupled with the field inversion technique to discover mathematical expressions from indirect observation data~\cite{he2024field}.
That is, the optimal Reynolds stress field is first inferred from indirect observation data with the adjoint-based method.
Further, the inferred field is used as training data to find an optimal symbolic expression based on the GEP method.
However, the GEP-based training strategy may result in relatively high computational costs, particularly when involving the RANS simulations in the training process~\cite{ZHAO2020}.
This is attributed to the extensive number of generations, often ranging in thousands, involved in the evolutionary process of candidate models.
Each candidate model needs to be evaluated in the RANS calculations for every generation, leading to a significant increase in computational costs.
Hence, it is necessary to develop efficient methods for learning symbolic turbulence models based on indirect observation data.

Given the merits of the neural network and symbolic regression, 
Lav et al.~\cite{LAV2023109140} proposed combining the two approaches to improve the predictive accuracy of learned symbolic turbulence models.
That is, the neural network is first used to construct complicated model functions due to its great expressive capability, and the symbolic regression is further used to transform the neural network-based functional mapping into simple mathematical expressions. 
Their work focuses on using direct observation data of the Reynolds stress for symbolic turbulence modeling. 
Learning from the indirect data by coupling the neural network and symbolic regression still lacks investigation, which is of practical interest for scenarios having only indirect observation.
Moreover, the feature importance analysis of the neural network is worthy of investigation to improve the efficiency of the symbolic regression by excluding unimportant features.

In this work, we propose a novel framework that combines neural networks and symbolic regression to learn interpretable turbulence models from indirect observation data.
Neural networks are employed to model the Reynolds stress, leveraging their great expressive power and training efficiency. 
The symbolic regression method~\cite{Silviu2020} further transforms the black-box neural network model into a white-box symbolic model composed of important flow features.
High-dimensional input features are frequently encountered in data-driven turbulence modeling~\cite{Fang2023}, often impeding the efficient discovery of accurate symbolic expressions.
To address this difficulty, the Permutation Feature Importance (PFI) method is introduced to identify and exclude insignificant, irrelevant, and redundant features, thereby simplifying the neural network and improving the efficiency of symbolic regression~\cite{Chen2017,Helali2024}.
We note that learning neural-network-based RANS models from indirect data~\cite{ling2016reynolds, singh2017machine} and converting neural network models to symbolic expressions ~\cite{LAV2023109140} both have been investigated.
Here, the combination of these two approaches is proposed, which provides a practical way to discover interpretable symbolic models from sparse, indirect data.
Moreover, the neural network is introduced as an intermediate process for symbolic modeling, which alleviates the efficiency issue of the conventional symbolic regression with the gradient-based optimization and the feature reduction.
The remainder of the paper is structured as follows.
In Section~\ref{sec:II}, we present the Reynolds stress representation, the neural-network training method, feature importance analysis, and the symbolic regression method.
In Section~\ref{sec:III}, the results of the proposed method are presented and discussed for flows in a square duct and flows over periodic hills.
Finally, the concluding remarks are provided in Section \ref{sec:IV}.

\section{Methodology}\label{sec:II}

\begin{figure}[!htb]
    \centering
    \includegraphics[width=\linewidth]{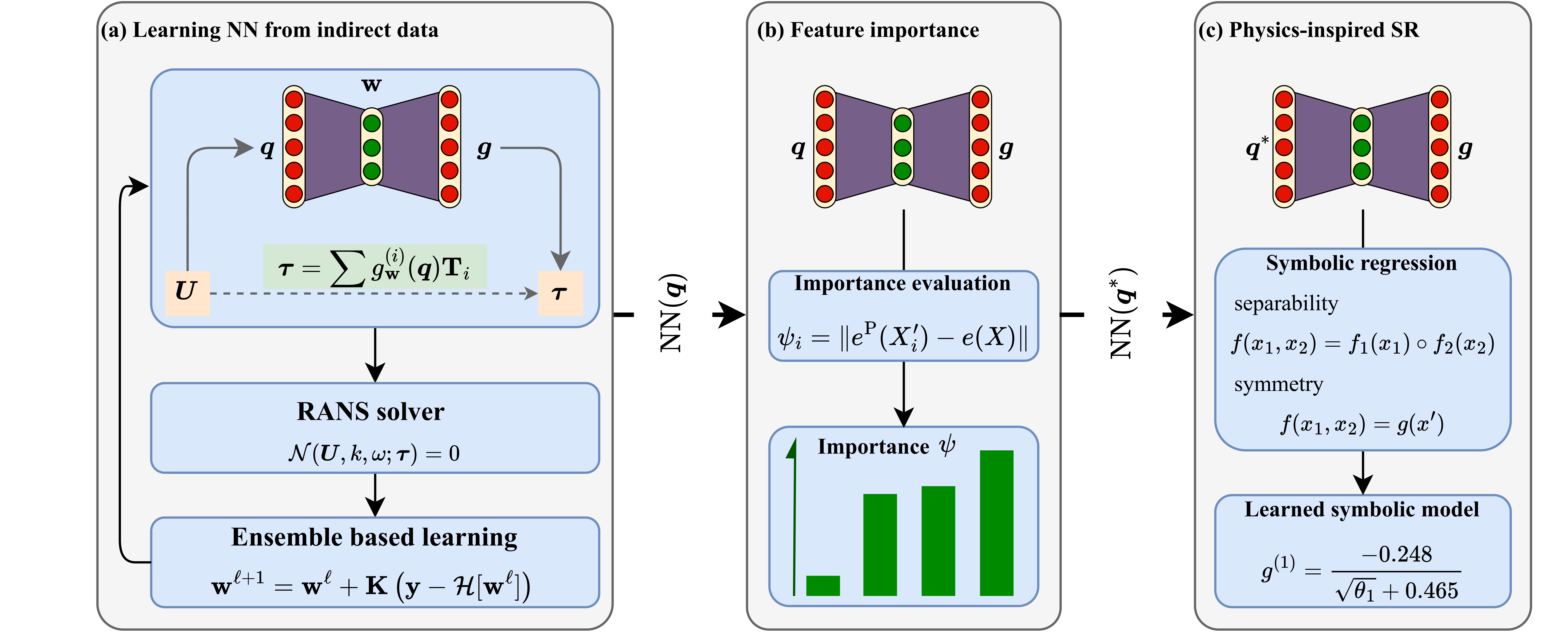}
    \caption{Schematic of the proposed neural network-based symbolic regression framework: (a)  neural network training using indirect data by the ensemble-based method, (b) feature importance evaluation using the permutation feature importance method, and (c) physics-inspired symbolic regression (i.e., AI Feynman) with important input features.}
    \label{fig:overall-floa-chart}
\end{figure}

We propose using the neural network associated with the feature importance analysis to learn symbolic turbulence models from indirect observation data. 
The neural network is used to approximate complicated functional mapping from indirect data, while the symbolic regression is further used to interpret the black-box network model to be a white-box mathematical expression. 
The proposed workflow for learning symbolic turbulence models from indirect data is presented in Figure~\ref{fig:overall-floa-chart}, which consists of the following three main steps.

\begin{enumerate}[label=(\roman*)]
    \item Neural network training. The neural network is used to represent the Reynolds stress~$\bm{\tau}$ with a prescribed set of input features~$\bm{q}$ based on the nonlinear eddy viscosity model.
    The ensemble Kalman method~\cite{iglesias2013ensemble} is used to minimize the discrepancy between the RANS prediction and indirect observation by optimizing the neural network weights~$w$.
    \item Feature importance analysis. The contribution portion~$\psi$ of the input features~$\bm{q}$ is identified with the feature analysis method, i.e., the permutation feature importance (PFI), in the trained neural network.
    In doing so, the insignificant features can be excluded to accelerate the following symbolic regression.
    \item Symbolic regression.
    The physics-inspired symbolic regression algorithm, i.e., AI Feynman~\cite{Silviu2020}, is employed to transform the learned neural network into a symbolic expression with identified important features~$\bm{q^{*}}$.
    The neural network outputs are used as the training data for the symbolic regression.
    The physical properties such as separability and symmetry can be used to divide the regression problem into simple ones, eventually leading to an explicit model formulation.
\end{enumerate}

Both the neural network-based training and the symbolic regression techniques are used in the proposed framework, which requires different model representations and training approaches. 
In the following, the representation method and the proposed training workflow are presented in detail, respectively.

\subsection{Neural network-based nonlinear eddy viscosity model}
\label{sec:tbnn}

Based on the generalized eddy viscosity assumption~\cite{pope_1975}, the anisotropic part of the Reynolds stress tensor, i.e., $\m{b}=\bm{\tau}/2k-\m{I}/3$, is assumed to have a functional dependency on the mean flow field as
\begin{equation}
\label{eq:b}
    \m{b}=\sum_{i=1}^{M}g^{(i)}\bra{\mathbf{\bm{\theta}}}\m{T}^{(i)} \text{,}
\end{equation}
where $\bm{\theta},\m{T}^{(i)}$ are scalar invariants and tensor bases respectively, and $g^{(i)}$ are tensor coefficient functions on scalar invariants~$\bm{\theta}$. 
A compact representation~\cite{Fu2011} with $6$ irreducible invariants and the first $5$ tensor bases is employed in this work.
These scalar invariants and tensor bases are
\begin{subequations}
    \begin{equation}
        \label{equ:invariants}
        \theta_1=\tr \{ \m{S}^2\},\; \theta_2=\tr \{\m{W}^2\},\; \theta_3=\tr \{\m{S}^3 \},\; \theta_4=\tr \{ \m{W}^2\m{S} \},\; \theta_5=\tr \{ \m{W}^2\m{S}^2 \},\; \theta_6=\tr \{\m{SW}\m{S}^2\m{W}^2 \} \text{,}
    \end{equation}
    \begin{equation}
        \label{equ:bases}
         \m{T}^{(1)}=\m{S},\; 
         \m{T}^{(2)}=\m{SW}-\m{WS},\;
         \m{T}^{(3)}=\m{S}^2-\tfrac{1}{3}\m{I}\tr \{\m{S}^2\},\;
         \m{T}^{(4)}=\m{W}^2-\tfrac{1}{3}\m{I}\tr \{\m{W}^2\},\;
         \m{T}^{(5)}=\m{W}\m{S}^2-\m{S}^2\m{W} \text{,}
    \end{equation}
\end{subequations}
where the tensors $\m{S},\m{W}$ are mean strain rate and rotation rate normalized by turbulence time scale $\mathcal{T}$ as $\m{S}=\mathcal{T}\tfrac{1}{2}\sbra{\nabla\bm{U}+\bra{\nabla\bm{U}}^{\top}}$ and $\m{W}=\mathcal{T}\tfrac{1}{2}\sbra{\nabla\bm{U}-\bra{\nabla\bm{U}}^{\top}}.$ 
For statistically two-dimensional flow, there are only the first 3 independent tensor bases~\citep{pope_2000}.
Moreover, the third tensor base $\m{T}^{(3)}$ can be absorbed into the pressure gradient term~\cite{STROFER2021TAML,zhang2022}, leaving the two tensor bases left.
Note that other flow features, such as the normalized eddy viscosity and the ratio of turbulence production to dissipation, can be also introduced as independent variables~$\bm{q}$ of the coefficient function~$\bm{g}(\bm{q})$ in practical data-augmented turbulence modeling \cite{Wang2017PhysRevFluids, Fang2023, LAV2023109140}. 
Here, the notation $\bm{q}$ denotes the augmented inputs of the tensor coefficient function $g^{(i)}$, which includes the scalar invariants $\theta_i$.

In this work, a feedforward neural network (FNN) is used to construct the mapping from input features $\bm{q}=[q_1,\cdots]^{\top}$ to tensor coefficients $\bm{g}=[g^{(1)},\cdots]^{\top}$, i.e.,
$\bm{g}=\mathcal{N}\bra{\bm{q};\m{w}}$,
where $\m{w}$ is the neural network weights.
The input features~$\bm{q}$ need to be scaled by
$\hat{q}_i=q_i/(1+\abs{q_i})$~\cite{Ling2015}.
In doing so, the scalar invariants can range within $[-1, 1]$, which enhances the robustness of the learned neural network model~\citep{Ling2015,Wu2017PhysRevFluids, Liu2023,Fang2023}.

The turbulence time scale can be estimated based on $\mathcal{T}=1/(C_\mu \omega)$, where $C_\mu = 0.09$ and $\omega$ is the specific dissipation rate that is obtained by solving the corresponding transport equation.
The $\omega$ equation can be given by $k$-$\omega$ SST turbulence model~\cite{menter2003ten} as presented in~\ref{sec:sst}.
Further, the nondimensional tensors $\m{S}$ and $\m{W}$ are used to construct the scalar invariants and other flow features at each time step during the RANS simulation.
These input features~$\bm{q}$ are propagated to the tensor coefficients~$\bm{g}$ through the neural networks and further the Reynolds stress anisotropy~$\m{b}$ based on Eq.~\eqref{eq:b}.
Finally, the obtained Reynolds stress is used to close the RANS equation and provide the prediction of mean flow fields.

\subsection{Ensemble Kalman method for learning neural network-based model from indirect data}

The neural network-based turbulence model can be trained with indirect observation data by involving the RANS calculations during the training process~\cite{zhang2022,zhang2023combining}. 
It amounts to optimize the neural network weights $\m{w}$ by minimizing the error metric between RANS calculations $\mathcal{H}\sbra{\m{w}}$ and indirect training data $\m{y}$, i.e.,
\begin{equation}
     \mathop{\arg\min} \limits_{\m{w}} J=\|\m{y}-\mathcal{H}\sbra{\m{w}}\| \text{,}
\end{equation}
where $\mathcal{H}$ is the model operator that maps the neural network weights to the mean flow predictions, and the operator $\|\Box\|$ denotes the $\mathrm{L}^2$ norm. In practice, the operator $\mathcal{H}$ combines the RANS solver with the subsequent post-processing steps to obtain the model prediction.
The ensemble Kalman method~\cite{strofer2021ensemble,zhang2022} is an ensemble-based statistical inference method, which is emerging to be used for learning neural network models~\cite{Kovachki_2019}.
 It uses sample statistics of model inputs and outputs to approximate the gradient of the cost function, which circumvents extra efforts to develop the adjoint solver in contrast to the conventional adjoint-based method~\cite{Duraisamy2021}.
 This is in stark contrast to the evolutionary optimization method that does not make use of the gradient information.
 The update scheme of the ensemble Kalman method can be derived based on the ensemble-based gradient and Hessian~\cite{luo2015iterative}.
 It can be formulated as 
\begin{equation}
     \m{w}_j^{\ell+1}=\m{w}_j^\ell+\m{K}\bra{\m{y}-\mathcal{H}\sbra{\m{w}_j^\ell}} \text{,}
\end{equation}
where $\ell$ is the iteration index, and $\m{K}$ is the Kalman gain matrix based on sample covariance between model inputs and outputs.
In practical implementation, the Kalman gain matrix is computed by~\cite{zhang2022}
     \begin{equation}
    \m{K}=\m{S}_\mathrm{w}\m{S}_\mathrm{y}^{\top}\bra{\m{S}_\mathrm{y}\m{S}_\mathrm{y}^{\top}+\m{R}}^{-1},
     \end{equation}
     where $\m{S}_\mathrm{w}$ and $\m{S}_\mathrm{y}$ are square-root matrices of neural network weights and model predictions.
    These matrices can be calculated by 
     \begin{subequations}
         \begin{equation}
             \m{S}^\ell_\mathrm{w}=\frac{1}{N-1}\sbra{\m{w}_1^\ell-\rey{\m{w}}^\ell,\cdots,\m{w}_N^\ell-\rey{\m{w}}^\ell},
         \end{equation}
         \begin{equation}
             \m{S}^\ell_\mathrm{y}=\frac{1}{N-1}\sbra{\mathcal{H}\sbra{\m{w}_1^\ell}-\rey{\mathcal{H}\sbra{\m{w}^\ell}},\cdots,\mathcal{H}\sbra{\m{w}_N^\ell}-\rey{\mathcal{H}\sbra{\m{w}^\ell}}},
         \end{equation}
     \end{subequations}
     where $N$ is the number of samples and the operator $\overline{\Box}$ indicates the sample mean.
    The detailed training procedure is illustrated in~\ref{sec:procedure}.
    The DAFI library~\cite{strofer2021dafi} is used to implement the ensemble-based Kalman method.
    We note that the present work focuses on steady-state applications with sparse data.
   However, the ensemble Kalman method may encounter sample collapse issues when the size of the observation (e.g., DNS data) and model variables (e.g., neural network weights) significantly exceed the number of samples.
   Limited samples can lead to spuriously large covariances between the model variables and observations, causing ensemble collapse~\cite{chen2010cross,zhang2022}. 
   Correlation-based localization~\cite{Luo2018} can be employed to mitigate this issue, allowing the ensemble Kalman method to effectively manage large-scale training data sets.

\subsection{Permutation feature importance}

Feature importance analysis is often used to enhance the interpretability of neural networks by providing the contribution value of each input feature to the model prediction~\cite{Wang2017PhysRevFluids,MANDLER2023}.
In this work, we leverage the feature importance analysis to facilitate the transformation from the black-box neural network model to a white-box symbolic model.
 The feature analysis method can inform important features for the subsequent symbolic regression, thereby simplifying the regression problem and improving the training efficiency.

Permutation feature importance (PFI) is one of the feature analysis methods, which can quantify the individual impact of input features on the statistical efficacy of a fitted model based on a designated tabular dataset.
 It is achieved by shuffling values of a single feature and subsequently assessing the consequent deterioration in the model performance~\cite{breiman_random_2001}.
 This approach can determine the extent to which the model depends on a specific feature by disrupting the intrinsic connection between the input features and the model prediction.
 The procedure of the PFI method is illustrated in Algorithm~\ref{alg:algorithm1}.
 The input consists of a trained neural network-based turbulence model $\phi$, a feature matrix $X$, and target data $Y$, where $X$ represents the dataset~{of input features} with each data point corresponding to a single CFD mesh cell, and $Y$ represents the output of model on data $X$.
 The model error $e$ is computed using the original feature matrix $X$ and target data $Y$. 
 Further, for each feature $q_i$, a permuted feature matrix $X_i'$ is generated by randomly shuffling the values of $q_i$ in $X$, and the error $e'_i$ is recalculated on the permuted data. 
 The feature importance for $q_i$ is determined by the difference in errors, $\psi_i=\|e'_i-e_i\|$, which quantifies the sensitivity of the model performance to the shuffling of each feature. 
 This process is repeated for all features to compute the overall importance ranking.

\begin{algorithm}[!htb]
	\caption{The permutation feature importance (PFI) algorithm.}
	\label{alg:algorithm1}
	\KwIn{Predictive model~$\phi$, tabular dataset including feature matrix $X$ and target data $Y$.}
	\KwOut{PFI values of each feature.}
	\BlankLine
	Estimate model error $e=\|\phi(X)-Y\|$ of the model $\phi$ on dataset $X$;
 
    \ForEach{feature $q_i \in \{q_1,q_2,\cdots\}$ }{
    Generate permuted feature matrix $X'_i$ by randomly shuffling the feature $q_i$ in the data $X$.

    Estimate model error $e'_i=\|\phi(X'_i)-Y\|$ on the permuted data $X'_i$.

    Obtain the PFI value for the feature $q_i$ as $\psi_i = \|e'_i-e\|$
    }
\end{algorithm}

\subsection{Physics-inspired symbolic regression}
The symbolic regression can be used to learn mathematical expressions consisting of identified important features, given data generated from the neural network model. 
For symbolic regression, functions can be represented as strings of symbols in reverse Polish notation to avoid the need for parentheses.
However, as the length of these strings increases, the number of possible combinations grows exponentially, making brute-force methods impractical.
 To address this challenge, biology-inspired strategies, e.g., genetic algorithm~\cite{dubvcakova2011eureqa}, are often used to search for optimal expression with mutation, selection, inheritance, and recombination.
 Various studies~\cite{WEATHERITT201622, ZHAO2020, LAV2023109140, Fang2023} have been conducted to use gene-expression programming (GEP) to find symbolic expressions for tensor coefficients.
 Besides, a physics-inspired strategy, namely AI Feynman~\cite{Silviu2020}, has been proposed to find mathematical expressions with a divide-and-conquer strategy.
 Specifically, this method recursively breaks hard problems into simple ones with fewer variables based on physical reasoning such as symmetry and separability in the data.
 It can be relatively efficient in contrast to the biology-inspired strategy~\cite{Silviu2020}.
 Therefore, we employ this method to learn symbolic models from the data generated with the neural network in this work.

The AI Feynman method assumes that functions $f(x_1,x_2,\cdots)$ encountered in physics frequently exhibit the following simplifying properties. 
 \begin{itemize}
     \item [(1)] Function $f$, or a portion thereof, is a low-degree polynomial. 
     This property facilitates polynomial fitting, allowing for swift resolution through the solution of a linear equation system to ascertain the polynomial coefficients.
     \item [(2)] Function $f$ is continuous on its domain, which facilitates the approximation of $f$ using a feed-forward neural network equipped with a smooth activation function.
     \item [(3)] Function $f$ exhibits translational or scaling symmetry concerning certain variables. 
     This facilitates the transformation of the regression problem into a simple one with fewer independent variables.
     \item [(4)] Function $f$ can be expressed as either a sum or a product of two parts, each devoid of any shared variables. 
     It allows for dividing independent variables into two separate sets, facilitating the transformation of the regression problem into two simple ones. 
 \end{itemize}
 
\begin{figure}[!htb]
     \centering
     {\includegraphics[width=0.75\linewidth]{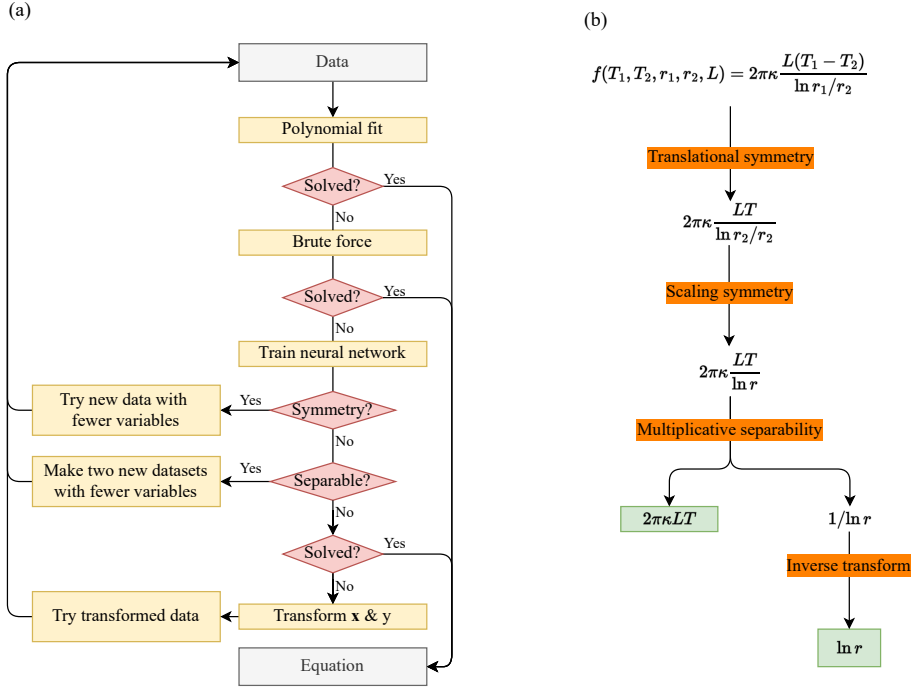}}
     \caption{(a) Flowchart of AI Feynman algorithm for symbolic regression. (b) An example of AI Feynman algorithm for learning symbolic expressions. 
     The AI Feynman method sequentially uses modules of polynomial fit, brute force, symmetry, and separability. 
     The entire problem can be broken down into simpler parts and recursively applies the full algorithm for each part once it cannot be solved.}
     \label{fig:AI-Feynman}
 \end{figure}

 The overall algorithm is illustrated in Figure~\ref{fig:AI-Feynman}(a).
 The method comprises a sequence of modules designed to leverage each of the aforementioned properties. 
 Once the entire problem cannot be solved, it seeks to divide and simplify it into manageable segments.
 Subsequently, the algorithm is recursively applied to each segment.
 Each module used in AI Feynman~\cite{Silviu2020} is illustrated briefly in the following.

 \begin{enumerate}[label=(\alph*)]
     \item \textbf{Polynomial fit.}
    This module is designed to test whether a problem can be solved using a low-order polynomial. 
    It employs the polynomial regression method to determine the polynomial coefficients that lead to the best agreement with data.

     \item \textbf{Brute force.} 
     This module is designed to find the symbolic expression for a problem that has been decomposed into simpler components by other modules.
     It systematically tests all possible symbolic expressions, representing them as strings of symbols using reverse Polish notation to avoid the need for parentheses.

     \item \textbf{Train neural network.}
     This module is designed to train a neural network to perform high-dimensional interpolation between available data points.
     With the trained neural network, we can evaluate $f$ at points $(x_1,x_2,\cdots,x_n)$ where data points are unavailable, thereby checking properties of symmetry and separability within data.
     The architecture of the neural network used in this work is recommended by the AI-Feynman code, which is a feed-forward, fully connected model with six hidden layers utilizing softplus activation functions.

     \item \textbf{Check symmetry.}
    The translational symmetry of a function is validated if $f(x_1,x_2,\cdots,x_n)=f(x_1+c,x_2+c,\cdots,x_n)$ for various constants $c$ within a specified precision.
    This implies that the function $f$ depends on the difference between $x_1$  and $x_2$.
    Consequently, the two variables can be reduced to one new variable $x'_1\equiv x_2-x_1$.
    Similarly, all pairs of input variables are tested to determine if any pair can be replaced by its sum (additive symmetry), product (multiplicative symmetry), or ratio (scaling symmetry).
    
     \item \textbf{Check separability.}
     A function is separable if it can be decomposed into two parts without shared variables.
     Both additive and multiplicative separability are tested.
     Taking the multiplicative separability as an example, it is assessed by computing the quantity
     \begin{equation*}
         \Delta_{\text{sep}}\equiv \abs{f(x_1,x_2)-\frac{f(x_1,c_2)f(c_1,x_2)}{f(c_1,c_2)}}
     \end{equation*}
     at each data point~$x$ for various constants $c$. The operator $\abs{\Box}$ represents the absolute value.
     If the average $\Delta_{\text{sep}}$ is less than a prescribed threshold, the function is considered multiplicatively separable.
     
     \item \textbf{Data transformation.}
     Once the problem is still not solved, various transformations are applied to the independent and dependent variable.
     These transformations may assist the brute force module in identifying underlying model formulations.
     The variable is transformed using the following functions: square root, square, logarithm, exponential, inverse, sine, cosine, arcsine, arccosine, and arctangent. 
     \end{enumerate}
     
 We present an example in Figure~\ref{fig:AI-Feynman}(b) to illustrate the procedure of the AI Feynman algorithm to discover a symbolic expression with symmetry and separability properties. The heat transfer rate of a pipe can be calculated as $2\pi \kappa L(T_1-T_2)/\ln(r_1/r_2)$ with length $L$, inner and outer radii $r_1,r_2$, and uniform wall temeratures~$T_1, T_2$. To find this symbolic expression, a neural network is first trained to fit given data, revealing translational symmetry, scaling symmetry, and multiplicative separability.
Further, the translational symmetry can eliminate one variable by defining $T=T_1-T_2$, while the scaling symmetry eliminates another variable by defining $r=r_2/r_1$.
The multiplicative separability allows for the factorization $\mathcal{F}(L,T,r)=G(L,T)H(r)$, splitting the problem into two simpler sub-problems. 
Finally, the function $G(L,T)$ can be solved $G(L,T)=2\pi\kappa LT$ using polynomial fitting, and the function $H(r)$ can be determined $H(r)=\ln r$ by brute force after applying an inverse data transformation.

\section{Test cases}\label{sec:III}

 We employ two cases to demonstrate the capability of the proposed method in finding symbolic turbulence models via neural networks associated with feature importance analysis.
The first case is secondary flow in a square duct, where we use synthetic data from RANS predictions with the Shih quadratic model~\cite{shih1993realizable}. 
This case aims to show the capability of the method to discover underlying symbolic expressions from indirect observation data. 
In the second case, the proposed method is applied with the DNS data for the separated flow over periodic hills, demonstrating the generalizability of the learned symbolic model to similar flow configurations.
In the following, the case details and results are presented, respectively.

\subsection{Flows in a square duct}

\subsubsection{Case setup}
 We first demonstrate the capability of the proposed method in discovering the underlying analytic model from indirect observation data for flows in a square duct.
 In the fully developed turbulent flow within a square duct,  secondary flows occur due to the imbalance between the Reynolds stress components, i.e., $\tau_{yy}-\tau_{zz}$.
 It has been demonstrated that linear eddy viscosity models are unable to predict the secondary flow \cite{SPEZIALE1982863,Speziale_1987}, which necessitates the use of nonlinear eddy viscosity models to capture the anisotropy of Reynolds stresses.
 Moreover, this flow case is inherently three-dimensional with six scalar invariants~$\bm{\theta}$, which can validate the capability of the PFI method to exclude insignificant features.
 We use velocity prediction from the RANS simulation with the Shih quadratic model \cite{shih1993realizable} as the training data.
 The Shih quadratic model provides mathematical expressions of the first four tensor coefficients~$\bm{g}$ based on the first two scalar invariants~$\bm{\theta}$ as
 \begin{equation}
     \begin{aligned}
         g^{(1)}(\theta_1,\theta_2)&=\frac{-2/3}{1.25+\sqrt{2\theta_1}+0.9\sqrt{-2\theta_2}},\\
         g^{(2)}(\theta_1)&=\frac{7.5}{1000+\bra{2\theta_1}^{3/2}},\\
         g^{(3)}(\theta_1)&=\frac{1.5}{1000+\bra{2\theta_1}^{3/2}},\\
         g^{(4)}(\theta_1)&=\frac{-9.5}{1000+\bra{2\theta_1}^{3/2}} \text{,}
     \end{aligned}
 \end{equation}
 where $\theta_1$ and $\theta_2$ are the invariants of nondimensionalized mean strain rate and rotation rate, as defined in Eq.\eqref{equ:invariants}.
In this case, we aim to show the capability of the proposed training strategy to discover the underlying symbolic expressions consistent with the true model formulation.

\begin{figure}[!htb]
     \centering
     \includegraphics[width=0.8\linewidth]{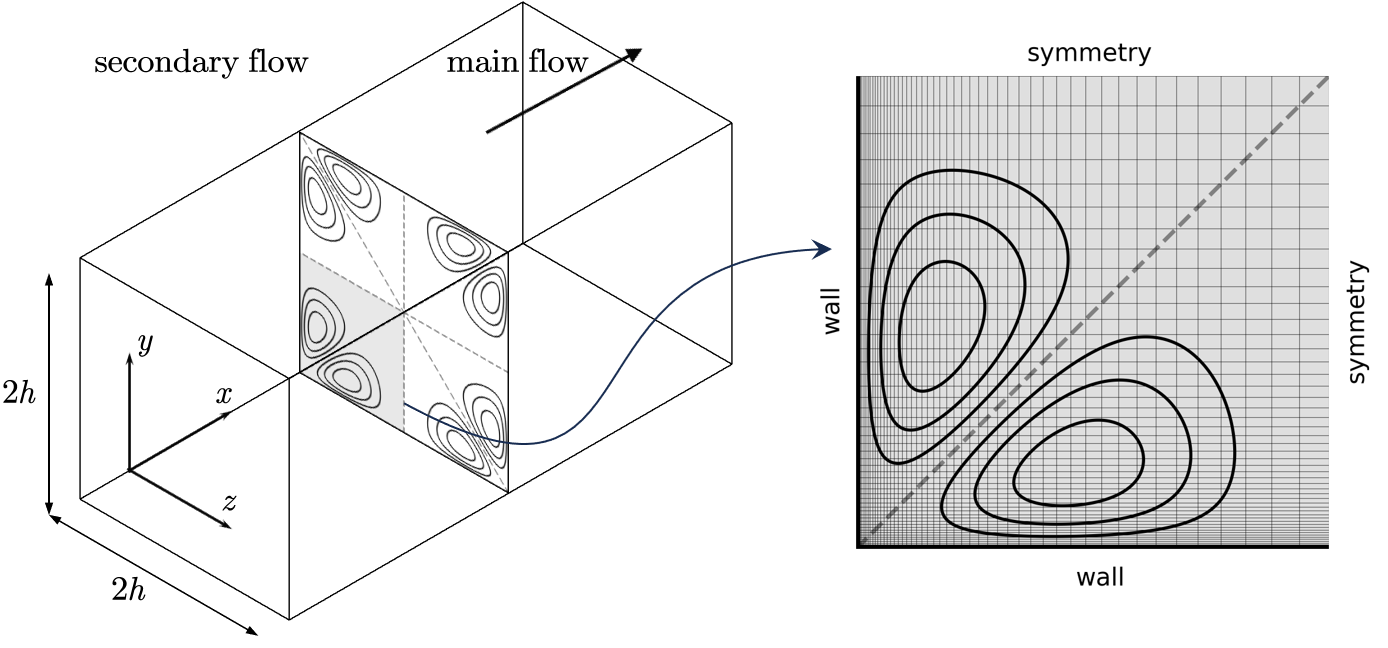} 
     \caption{Flow configuration and computational mesh for flows in a square duct. 
     The main flow aligns with the $x$ direction, and the secondary flow occurs within the $y-z$ cross-section due to the Reynolds stress imbalance. 
     The two-dimensional mesh covers a quarter of the cross-section. 
     The no-slip and symmetry conditions are applied to the boundaries, respectively.}
     \label{fig:duct-flow-config}
 \end{figure}

The flow is fully developed and statistically homogeneous in the streamwise direction, allowing the use of a two-dimensional mesh in the cross-sectional plane, as illustrated in Figure~\ref{fig:duct-flow-config}.
The Reynolds number based on the bulk velocity and the half-width of the duct is $Re_h=\num{3500}$.
The computational domain encompasses a quarter of the cross-section, with no-slip wall and symmetry boundary conditions applied.
 The mesh grid consists of $50 \times 50$ cells, refined near the solid walls.
 The turbulent kinetic energy $k$ and dissipation rate~$\varepsilon$ from the $k-\varepsilon$ model~\cite{LAUNDER1974} are used to provide the turbulence time scale $\mathcal{T}=k/\varepsilon$ for feature normalization in this case.

The three velocity components predicted by the RANS method with the Shih quadratic model in all the mesh cells are used as training data.
 The neural network architecture includes an input layer with six neurons, corresponding to the scalar invariant~$\theta_i$, two hidden layers with 10 neurons each, and an output layer with five neurons, representing each tensor basis coefficient~$g^{(i)}$.
 The activation function used is ReLU in this case. 
A total of $50$ samples are utilized for neural network training, based on our previous sensitivity study~\cite{zhang2022}.

As for the setup of the symbolic regression, the time limit for each brute-force search is set to $120$ seconds.
The maximum degree of the polynomial fit is set to $2$.
The number of epochs for the neural network training is set to $1600$.
The functions utilized in the brute force process are listed in Table~\ref{tab:AI-func}.
The unary function $\texttt{>}$ and $\texttt{<}$ increase and decrease a value by $1$, respectively. 
The function $\texttt{$\sim$}$ returns the negation of a value. 
The function $\texttt{I}$ computes the reciprocal, while $\texttt{A}$ and $\texttt{R}$ represent the absolute value and square root functions, respectively.

\begin{table}[!htbp]
  \centering
  \caption{The fundamental functions explored by the brute-force search in AI Feynman~\cite{Silviu2020}.}
  \resizebox{\textwidth}{!}{
    \begin{tabular}{lccc|ccccccc|cccc}
    \toprule
          & \multicolumn{3}{c}{nonary} & \multicolumn{7}{c}{unary}  & \multicolumn{4}{c}{binary} \\
    \midrule
    symbol & \texttt{0} & \texttt{1}    & \texttt{p}     & \texttt{>}     & \texttt{<}  & \texttt{$\sim$}   & \texttt{I}    & \texttt{L}    & \texttt{A}    & \texttt{R}    & \texttt{+}    & \texttt{*}    & \texttt{-}    & \texttt{D} \\
    meaning & $0$     & $1$     & $\pi$    & {\footnotesize increment} & {\footnotesize decrement} & {\footnotesize negative} & {\footnotesize invert} & {\footnotesize logarithm} & {\footnotesize abs}   & {\footnotesize sqrt}  & {\footnotesize add}   & {\footnotesize multiply} & {\footnotesize subtract} & {\footnotesize divide} \\
    \bottomrule
    \end{tabular}
    }
  \label{tab:AI-func}
\end{table}

\subsubsection{Results}
\begin{figure}[!htb]
        \centering
        \includegraphics[]{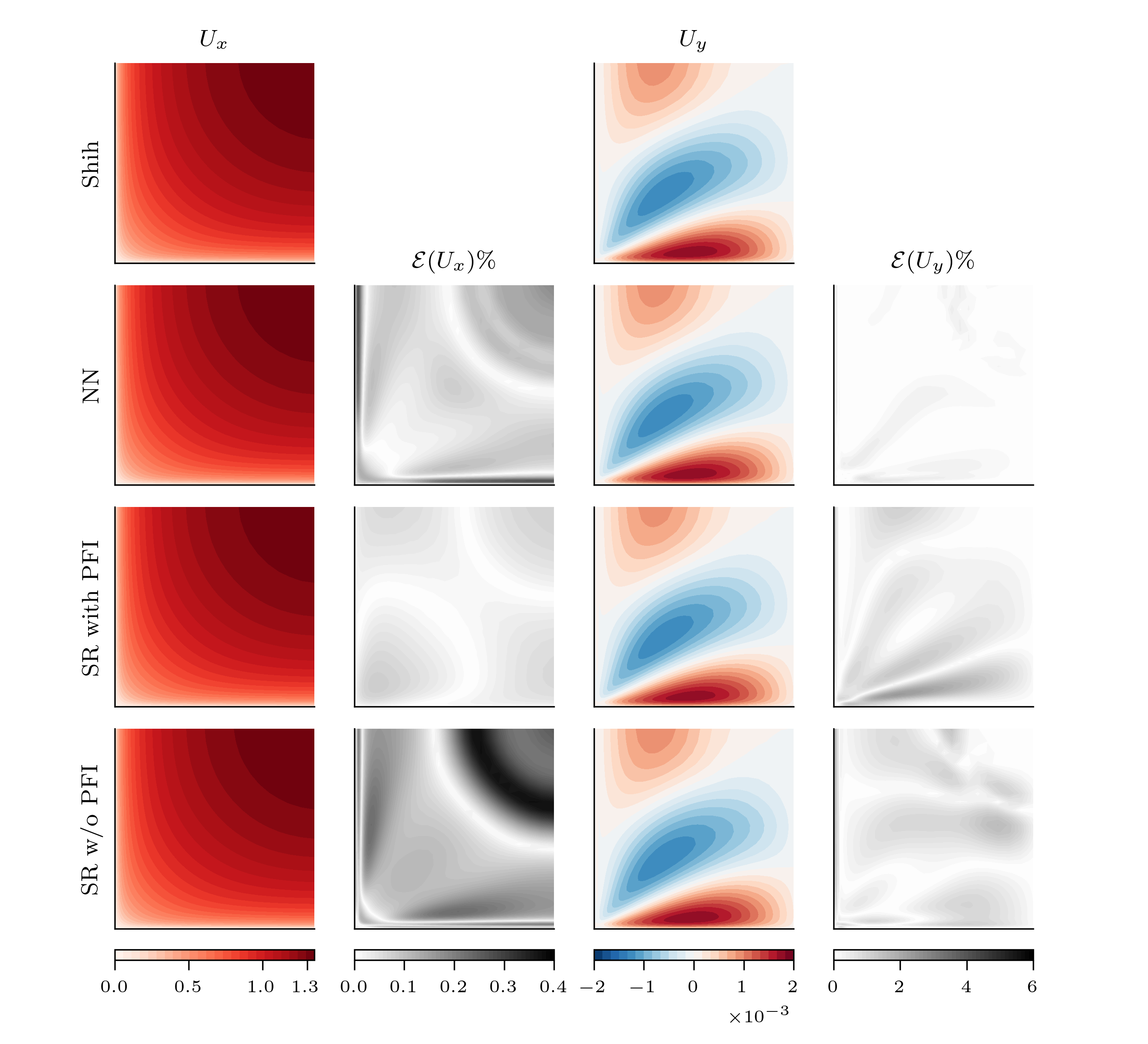}
        \caption{Velocity predictions of the neural network (NN) model, the symbolic regression (SR) model with PFI, the SR model without PFI, and the synthetic truth (Shih model) for the square duct case. 
        The four columns show the velocity component $U_x$, the error of $U_x$, the velocity component $U_y$, and the error of $U_y$, respectively. 
        The error is calculated by $\mathcal{E}(U_i)=\|U_i-U_i^{\text{truth}}\|/\max\bra{U_i^{\text{truth}}} $. 
        }
        \label{fig:velocity-comparison-square-duct}
    \end{figure}

The velocity fields predicted by the learned turbulence models based on the neural network, the symbolic regression with the PFI analysis, and the symbolic regression without the PFI analysis are presented in Figure~\ref{fig:velocity-comparison-square-duct}.
 Given the symmetry of the flow, only the velocity components $U_x$ and $U_y$ are plotted. 
 It can be seen that the learned neural network model can accurately predict the velocity field in both the streamwise flow and the in-plane secondary flow. 
 The streamwise velocity $U_x$ is influenced by the Reynolds stress component $\tau_{xy}$ and $\tau_{xz}$, while the in-plane velocities $U_y$ and $U_z$ are influenced by the Reynolds stress components $\tau_{yz}$ and $\tau_{yy}-\tau_{zz}$~\cite{michelen2021machine}, which will be presented and discussed in the following.
 A slight increase in the error of the streamwise velocity occurs in the center of the duct, where the scalar invariants $\bm{\theta}$ are close to zero.
 This is likely due to the significant discrepancy between the symbolic models with and without the PFI analysis in the range of small scalar invariants.
 The symbolic model with the PFI analysis can produce similar velocity predictions to the neural network model, indicating the effectiveness of the proposed training strategy.
 Except in the very near wall region, the symbolic regression models, with and without PFI, show no significant decline in prediction accuracy compared to the neural network model.

\begin{figure}[!htb]
        \centering
        \includegraphics[]{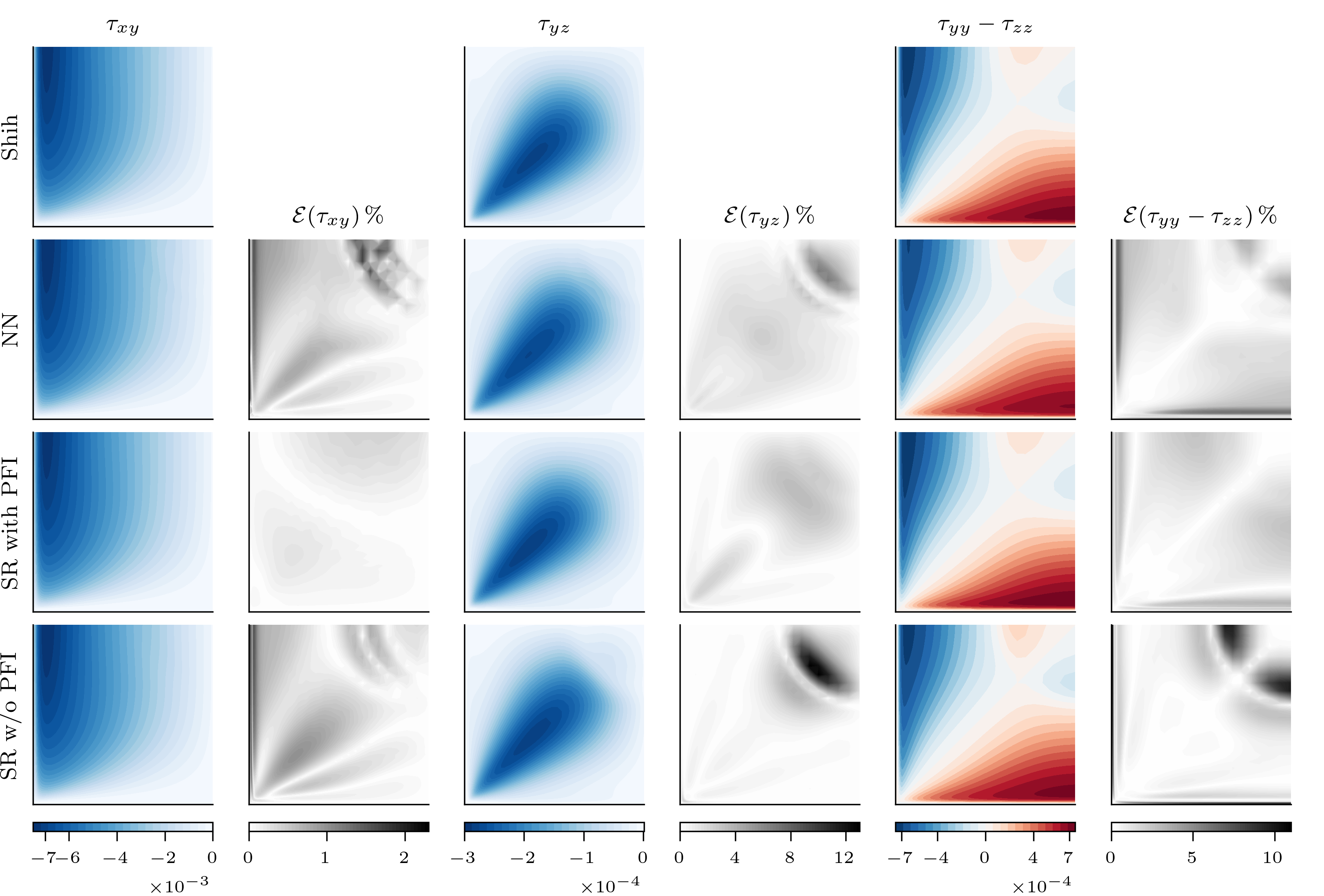}
        \caption{Reynolds stress predictions of the neural network (NN) model, the symbolic regression (SR) model with the PFI analysis, the SR model without the PFI analysis, and the synthetic truth (Shih model) for the square duct case. 
        The six columns show the shear stress $\tau_{xy}$, the error of $\tau_{xy}$, the shear stress $\tau_{yz}$, the error of $\tau_{yz}$, the normal stresses imbalance $\tau_{yy}-\tau_{zz}$, and the error of $\tau_{yy}-\tau_{zz}$, respectively. 
        The error is calculated by $\mathcal{E}(\tau_{ij})=\|\tau_{ij}-\tau_{ij}^{\text{truth}}\|/\max\bra{\tau_{ij}^{\text{truth}}} $. 
        }
        \label{fig:Reynolds-comparison-square-duct}
    \end{figure}

We compare the learned models and Shih quadratic model in the Reynolds stresses of $\tau_{xy},\tau_{yz},\tau_{yy}-\tau_{zz}$, as shown in Figure~\ref{fig:Reynolds-comparison-square-duct}.
The plot of Reynolds stress~$\tau_{xz}$ is omitted due to the flow symmetry.
The symbolic model with the PFI analysis exhibits the least prediction error for the Reynolds stress, while the symbolic model without the PFI leads to the largest prediction error. 
Moreover, the symbolic model with the PFI reduces the prediction errors of $\tau_{xy}$ across the entire cross-sectional plane and $\tau_{yy}-\tau_{zz}$ near the wall. 
 In contrast, the symbolic model without the PFI increases the prediction errors of $\tau_{yz}$ and $\tau_{yy}-\tau_{zz}$ near the center of the cross-sectional plane.
 The error contours are also provided in Figure~\ref{fig:Reynolds-comparison-square-duct}, which show 
that all learned models are able to improve the predictive accuracy of the Reynolds stress. 
However, it is noticeable that the symbolic model without the PFI exhibits relatively inferior prediction in both velocity and Reynolds stress in this case. 
It demonstrates that the feature importance analysis can enhance the predictive accuracy of the learned symbolic model.
    
\begin{figure}[!htb]
    \centering
    \includegraphics{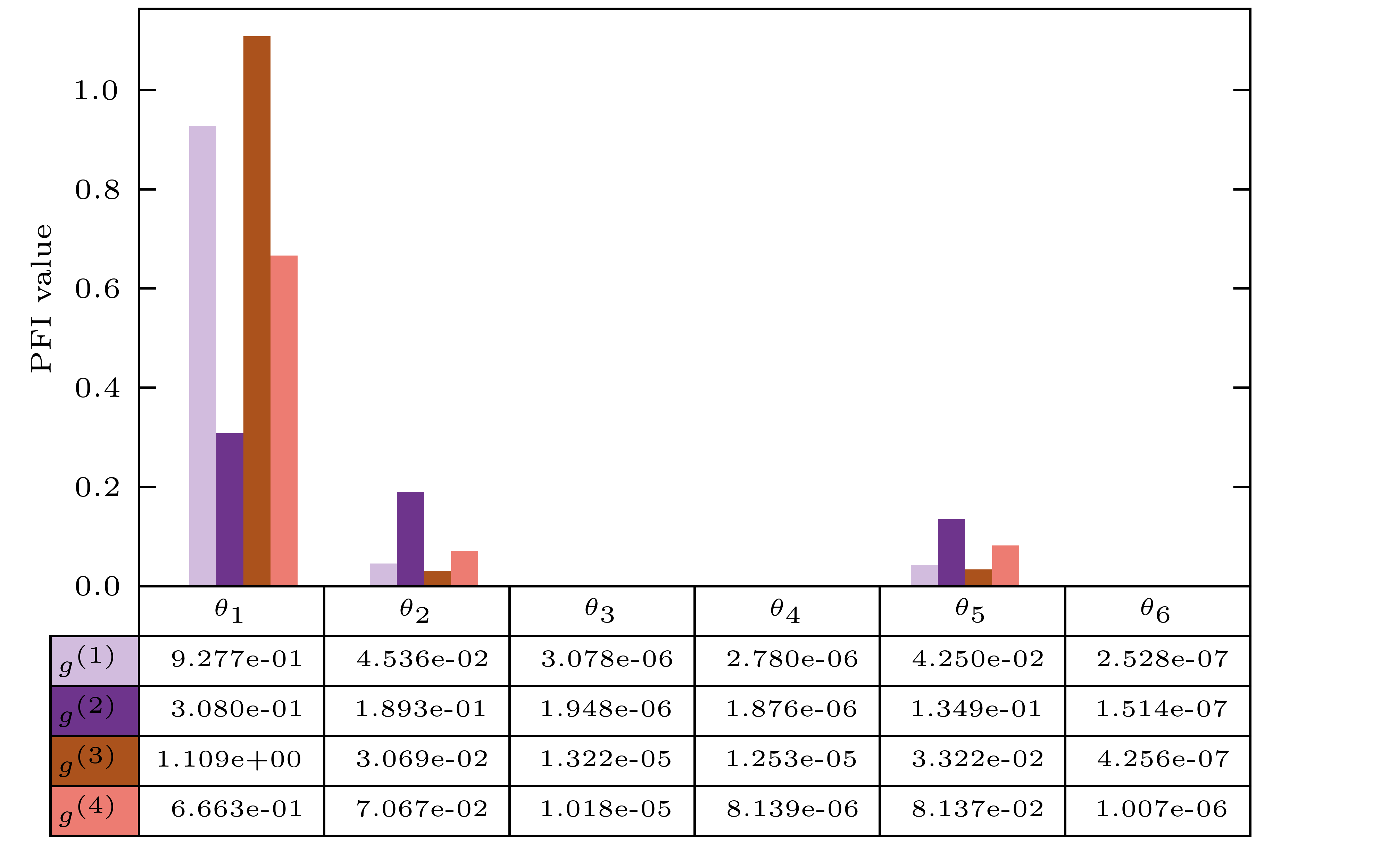}
    \caption{The PFI analysis for the six model input features $\theta_{1\sim 6}$ of the neural network model for the square duct case. 
    }
    \label{fig:PFI-value-Square-duct}
\end{figure}

The PFI values of the learned model are shown in Figure~\ref{fig:PFI-value-Square-duct}, which is consistent with the true model, i.e., Shih quadratic model.
 Specifically, the features of $\theta_3, \theta_4, \theta_6$ exhibit nearly zero PFI values, indicating that the models are nearly independent of these features. 
 The input feature $\theta_1$ has the largest PFI value, which is approximately one order of magnitude larger than those of $\theta_2$ and $\theta_5$.
Moreover, the input features $\theta_2$ and $\theta_5$ have the PFI values at similar magnitudes.
The notable PFI values of features $\theta_2$ and $\theta_5$  for all coefficient functions $g^{(i)}$ may be due to their dependence on feature $\theta_1$.
Specifically, the streamwise velocity component is orders of magnitude larger than the in-plane velocities \cite{michelen2021machine}, highlighting the dominance of the streamwise flow. 
This dominance leads to the approximate relationships among the invariants as $\theta_2\approx -\theta_1$ and $\theta_5\approx -\theta_1^2/2$.
 The PFI analysis indicates that $\theta_1$, $\theta_2$, and $\theta_5$ are relatively important features among the model inputs, while $\theta_3$, $\theta_4$, and $\theta_6$ can be excluded for symbolic regression.

\begin{figure}[!htb]
        \centering
        \includegraphics[]{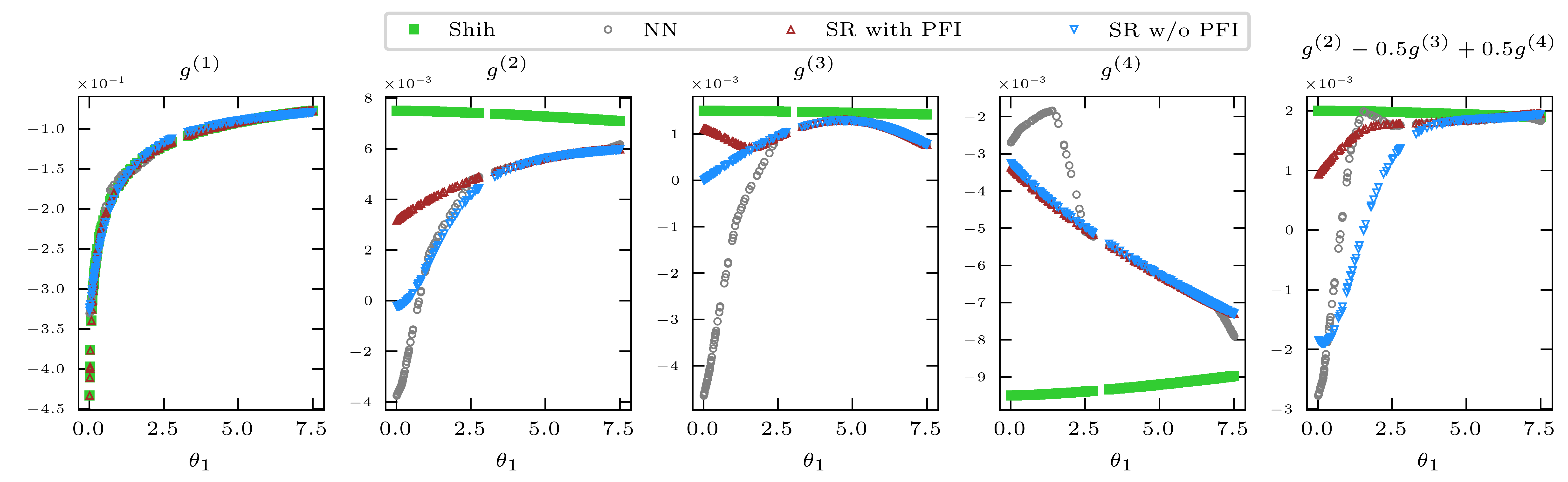}
        \caption{Comparison of tensor basis coefficients $g^{(1)},g^{(2)},g^{(3)},g^{(4)}$ and the combination $g^{(2)}-0.5g^{(3)}+0.5g^{(4)}$ with respect to the scalar invariant $\theta_1$ among the Shih model, neural network (NN) model, and symbolic regression (SR) models with and without the PFI analysis. }
        \label{fig:coefficients-on-theta1}
    \end{figure}

With dependent and unimportant features excluded, the proposed method can learn a symbolic expression consistent with the true model for the first tensor coefficient $g^{(1)}$.
We choose the data ranging of $\theta_1>2.5$, as the representation of small scalar invariants is limited, which would lead to incorrect functional mapping~\cite{zhang2022}. 
The obtained formula has a similar mathematical form to Shih's quadratic model as 
    \begin{equation}
        \label{equ:duct-1}
        \begin{aligned}
        g^{(1)}&=\frac{-0.248}{\sqrt{\theta_1}+0.465} \text{,} \\
        g^{(1)}_\text{Shih}&=\frac{-0.238}{\sqrt{\theta_1}+0.442} \text{,}
        \end{aligned}
    \end{equation}
    where Shih's model is reformulated based on $\theta_2 \approx - \theta_1$.
    However, the other three coefficient functions of $g^{(2-4)}$ do not agree with the true model, which is expressed as
    \begin{equation}
        \label{equ:duct-2}
        \begin{aligned}
        g^{(2)}&=\num{3.157e-3}+\frac{1}{1000}\frac{\theta_1}{\left(\sqrt{e^{\theta_1}}+\theta_1\right)^{1/4}},\\
        g^{(3)}&=\num{0.390e-3}+\frac{1}{1000}\log\left(1+\sqrt{-2+\pi-\sin\theta_1}\right),\\
        g^{(4)}&=-\num{2.367e-3} \sqrt{\theta_1 + 2} \text{.}
        \end{aligned}
    \end{equation}
    That can be caused by the ill-posedness of inferring the symbolic expressions from indirect observation data.
    On the one hand, the neural network is learned with the ensemble-based gradient approximation, which can have non-negligible training errors.
    Such errors could be further enlarged in the symbolic regression, leading to inconsistent model formulations.
    On the other hand, different model forms can provide similar velocity predictions in regions with small velocity gradients, 
    since the scalar invariants~$\theta_i$ may have little effect on the Reynolds stress and further the velocity prediction~\cite{Luo2024}. 
    Besides, indirect velocity data can only inform the combination of tensor coefficients~$g^{(2-4)}$ as shown in Figure~\ref{fig:coefficients-on-theta1}, i.e., $g^{(2)}-0.5g^{(3)}+0.5g^{(4)}$, as this quantity has effects on the in-plane velocity predictions~\cite{STROFER2021TAML}.
    For these reasons, the underlying model formulations of tensor coefficients~$g^{(2-4)}$ are challenging to learn from the indirect velocity data. Admittedly, the symbolic model in Eq.\eqref{equ:duct-2} is not informative physically, while it is relatively interpretable than the black-box neural network due to the explicit model expression.
    
    The feature importance analysis method could provide relatively simple model forms with good predictive accuracy.
    It is supported by the comparison between the learned symbolic models with and without the PFI analysis as shown in Figure~\ref{fig:coefficients-on-theta1}.
    Specifically, symbolic expressions using all six scalar invariants can be inferred, which does not yield the target model. 
    The obtained expressions involving six input features are written as
    \begin{equation}
        \label{equ:duct-6}
        \begin{aligned}
            g^{(1)}&=\num{-5.627e-2} \exp\bra{\frac{\pi }{\theta _1+\sqrt{\pi }}},\\
            g^{(2)}&=\num{-7.190e-3} \tan ^{-1}\left(\theta _1+\theta _2+\theta _3+\tfrac{\pi}{4}   \left(\theta _5-1\right)\right)-0.005,\\
            g^{(3)}&=\num{-1.293e-3} \sin \left(\frac{\theta _2}{\cos \left(\theta _3-\theta _6\right)+2}\right),\\
            g^{(4)}&=\num{-2.388e-3} \sqrt{-\theta _2+\sin \left(e^{\theta _4}\right)+1} \text{.}
        \end{aligned}
    \end{equation}
    The obtained model form is relatively complicated compared to the learned symbolic model with the PFI analysis.
    Moreover, the predictive accuracy of the symbolic model without the PFI analysis is larger than that of the symbolic model with the PFI analysis, compared to the synthetic truth.
    
The errors of the learned models with the neural network and the symbolic regression with and without the PFI analysis are summarized in Table~\ref{tab:duct-sr}.
It quantitatively shows that the PFI analysis can improve the accuracy of the learned symbolic turbulence model.
    Specifically, the symbolic model with the PFI leads to the error of $7.295 \times 10^{-5}$ in the predicted tensor coefficient $g^{(1)}$, while the symbolic model without the PFI provides a relatively large error at $5.522 \times 10^{-3}$.
    On the other hand, the symbolic regression without the PFI needs longer training times compared to using the PFI analysis as presented in Table~\ref{tab:duct-sr}.
    That is, the average computational time for Eq.\eqref{equ:duct-1} is $\SI{176}{min}$, while that of Eq.\eqref{equ:duct-6} is $\SI{1237}{min}$.
    
 \begin{table}[!htbp]
      \centering
      \caption{Summary of the prior fitting error, the posterior prediction error, and the computational cost. 
      The prior fitting error is calculated based on the sample mean of $\|g^{(i)}-g^{(i)}_\text{Shih}\|/\max\|g^{(i)}_\text{Shih}\|$. 
      The posterior prediction error is calculated based on the sample mean of $\|\bm{U}-\bm{U}_\text{Shih}\|/\max\|\bm{U}_\text{Shih}\|$.}
        \begin{tabular}{lcccccc}
        \toprule
            Model    & $g^{(1)}$    & $g^{(2)}$    & $g^{(3)}$    & $g^{(4)}$ & $U$ & CPU time\\
        \midrule
         NN    & \num{4.713e-03}     & \num{3.428e-01}     & \num{6.857e-01}     & \num{3.243e-01} & {\num{6.253e-04}} & \\
              SR with PFI & \num{7.295e-05}     & \num{2.385e-01}     & \num{2.756e-01}     & \num{3.033e-01} & {\num{3.138e-04}} & \SI{176}{min} \\
        SR w/o PFI & \num{5.522e-03}     & \num{3.088e-01}    & \num{3.508e-01}    & \num{3.049e-01} & {\num{9.934e-04}} & \SI{1237}{min} \\
        \bottomrule
        \end{tabular}
      \label{tab:duct-sr}
    \end{table}

\subsection{Flows over periodic hills}

\subsubsection{Case setup}
The flow over periodic hills is a canonical case characterized by flow separation and reattachment phenomena, which are often challenging to accurately predict using conventional linear eddy viscosity turbulence models~\cite{almeida1993}.
 Here a neural network-based model is first trained using the DNS data \cite{XIAO2020104431} and further transformed into symbolic form based on the feature importance analysis and symbolic regression.

\begin{figure}[!htb]
        \centering
        \includegraphics[width=0.8\linewidth]{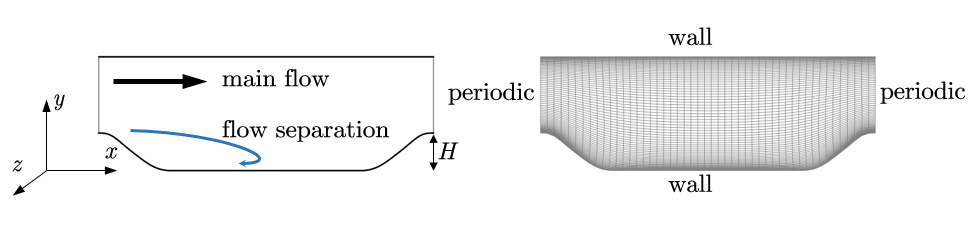}
        \caption{
        Schematic diagram of flows over the periodic hills. 
        The primary flow direction is along the $x$-axis, with flow separation occurring on the leeward side. 
        The height of the hills is denoted by $H$.
        Wall boundary conditions are applied to the upper and lower surfaces, and periodic boundary conditions are applied to the left and right sides.}
        \label{fig:config-phill}
\end{figure}

The geometry of the periodic hills along with the mesh is shown in Figure~\ref{fig:config-phill}.
 Since the flow is statistically homogeneous in the spanwise direction, a 2-dimensional mesh is used.
 The size of the mesh is $100\times 150$ in the stream-wise and wall-normal directions respectively and is refined near the wall.
 A periodic boundary condition is applied on the inlet and outlet of the computational domain, and no-slip wall boundary conditions are imposed on the bottom and top surfaces.
 The Reynolds number based on the bulk velocity and hill height is $Re_H=5600$.

For statistically two-dimensional turbulent flow, there only exists the first two invariants and the first two tensor bases \cite{pope_2000,zhang2022}.
 We augment the input features with three other flow quantities, i.e., normalized eddy viscosity, the ratio of turbulent kinetic energy production to dissipation, and the ratio of total to isotropic part of Reynolds stress. 
 The introduced three features are Galilean invariance and are often used in data-driven turbulence modeling~\cite{Ling2015, Wang2017PhysRevFluids, HE2022, Fang2023}. 
 The description of the five input features is listed in Table~\ref{tab:features}. 
 The turbulent kinetic energy $k$ and dissipation rate $\omega$ are modeled by the $k$-$\omega$ SST model in this case. 
 The Python code for the proposed method and the test case of the flows over periodic hills are provided in a publicly available GitHub repository~\cite{renkf-git}.
 
\begin{table}[!htb]
    \centering
    \begin{tabular}{ccl}
    \toprule
    feature & physical quantity & description \\
    \midrule
     $q_1$ & $\tr\{\m{S}^2\}$ &  Norm of nondimensionalized strain rate tensor\\
     $q_2$ & $\tr\{\m{W}^2\}$ &  Norm of nondimensionalized rotational rate tensor\\ 
     $q_3$ & $\nu_t/\nu$ &  Ratio of eddy viscosity to molecular viscosity\\
     $q_4$ & $\tilde{\mathcal{P}}/(\beta^*\omega k)$ &  Ratio of turbulent kinetic energy production to dissipation \\
     $q_5$ & $\|\bm{\tau}\|/k$ &  Ratio of total to isotropic part of Reynolds stresses\\
    \bottomrule
    \end{tabular}
    \caption{Non-dimensional flow features as neural network inputs for flows over periodic hills.}
    \label{tab:features}
\end{table}

The stream-wise and wall-normal velocity components at positions $x/H=0,1,3,5,7$ of DNS results are used as training data.
We draw 16 samples to learn a neural network-based model with the ensemble Kalman method.
The sample size for the case of the periodic hills has been systematically investigated in our previous study~\cite{Luo2024}, suggesting a sample size between 10 and 40 is optimal for ensemble Kalman-based training of neural networks.
Increasing the sample size beyond this range does not necessarily improve prediction accuracy. 
Given this, a sample size of 16 is used in this work to achieve a good compromise between computational efficiency and training accuracy.
The neural network has 5 inputs, 2 outputs, and 2 hidden layers with 10 neurons.
 The activation function is ReLU.
 The relative standard deviation of the neural network weights is set to $0.001$.
 For the observation data, the relative standard deviation is set to \num[]{1.e-6}.
 To investigate the robustness of the training process to data noise, we add different levels of random noise to the training data.
 The detailed results of this robustness study in terms of data noise are presented in~\ref{sec:noise-check}. 
The neural network training can accommodate training data noise with a relative standard deviation of up to $0.01$.
 These parameters, along with the number of samples, are determined based on our sensitivity study to have an optimal balance between training accuracy and efficiency.
 The setup for symbolic regression is consistent with the square duct case.

\subsubsection{Training results}

\begin{figure}[!htb]
        \centering
        \includegraphics[]{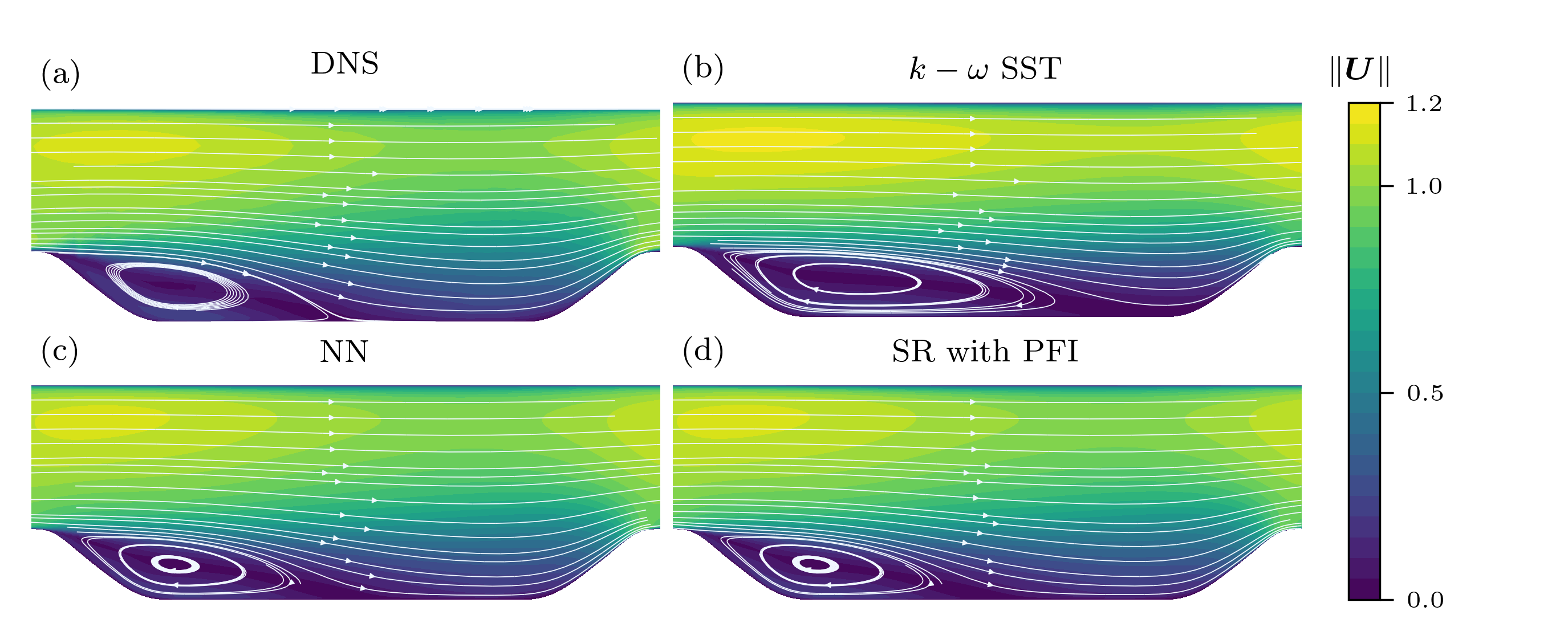}
        \caption{Comparison of streamline and velocity contour among (a) DNS, (b) the $k$-$\omega$ SST, (c) the neural network (NN) model, and (d) the symbolic regression (SR) model with the PFI analysis.}
        \label{fig:phill-streamline}
\end{figure}

The velocity contours are presented in Figure \ref{fig:phill-streamline}, which compares the streamlines of flow predicted by $k$-$\omega$ SST, DNS, and the learned models with the neural network and the proposed method.
 It can be seen from the contour plots that the baseline $k$-$\omega$ SST model overestimates the separation bubble size compared to the DNS results.
In contrast, both the symbolic model and neural network model can well predict the size of the separation bubble in good agreement with the DNS data without significant discrepancy. 
The symbolic regression without the PFI analysis learns a complicated model formulation that leads to divergence in the posterior prediction. 
Hence, only the results of the symbolic regression with the PFI are plotted here. 
The formulations of the learned symbolic models with and without the PFI analysis are presented in~\ref{sec:all-symbol}. 

\begin{figure}[!htb]
        \centering
        \includegraphics[]{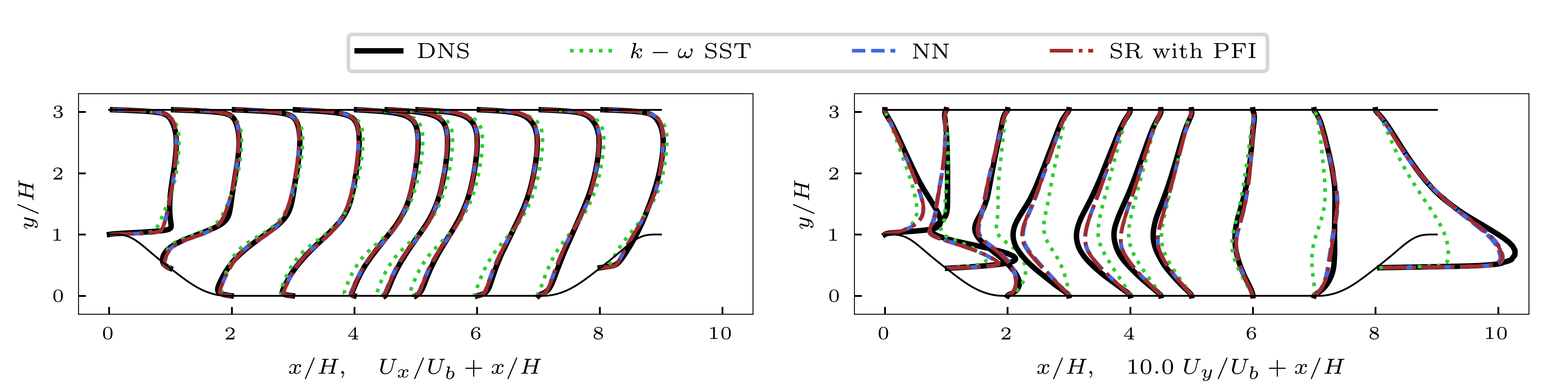}
        \caption{Comparison of the mean velocity profiles at $x/H=0,1,2,3,4,4.5,5,6,7,8$ in the (a) stream-wise and (b) wall-normal directions among the baseline model ($k$-$\omega$ SST), the DNS results, the neural network (NN) model, and the symbolic regression (SR) model.
        }
        \label{fig:phill-profile}
\end{figure}

The quantitative comparison of the velocity prediction is provided in Figure~\ref{fig:phill-profile}, where velocity profiles at positions $x/H=0,1,2,3,4,5,6,7,8$ are plotted.
 It can be seen that the symbolic model improves the predictions of velocity profiles in the separation bubble and reattachment region, where the baseline model fails to predict accurately.
 Also, the velocity prediction can be improved not only at the observed positions but also at the unobserved positions with the learned symbolic model. 
 
The prediction error of the learned models and the computational cost of the symbolic regression with and without PFI are summarized in Table~\ref{tab:addlabel}. 
It can be seen that the symbolic model learned with the PFI analysis can achieve a lower posterior error at $4.432 \times 10^{-2}$ compared to the neural network model at $4.407 \times 10^{-2}$.
However, the symbolic model learned without the PFI analysis leads to the divergence of the RANS solver.
On the other hand, the symbolic regression with the PFI analysis is more efficient than that without the PFI.
Specifically, the symbolic regression with the PFI can converge at around $\SI{123}{min}$, while the symbolic regression without the PFI requires $\SI{818}{min}$ for a converged result.

\begin{table}[!htb]
  \centering
  \caption{Prediction error in velocity and computational costs for symbolic regression with and without the PFI analysis for the periodic hill case.}
    \begin{tabular}{clc}
    \toprule
    \multirow{3}[2]{*}{Prediction error $\mathcal{E}(U)$} & NN    & \num{4.432e-02} \\
          & SR with PFI & \num{4.407e-02} \\
          & SR without PFI & Diverge \\
    \midrule
    \multirow{2}[2]{*}{Computational cost} & SR with PFI & \SI{123}{min} \\
          & SR without PFI & \SI{818}{min}\\
    \bottomrule
    \end{tabular}
  \label{tab:addlabel}
\end{table}

\begin{figure}[!htb]
        \centering
        \includegraphics{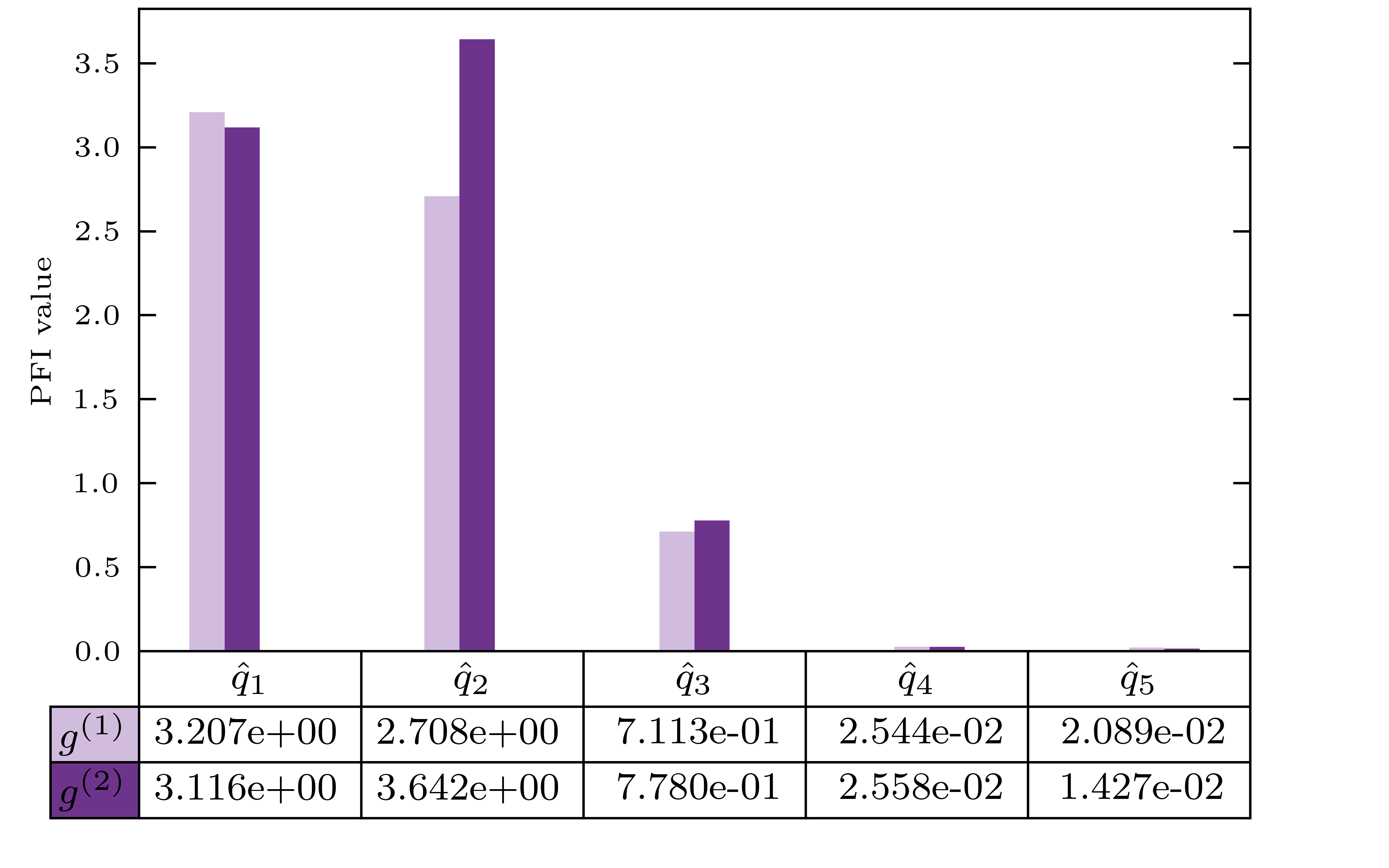}
        \caption{Plots of the PFI values for the five model input features $\hat{q}_{(1 \sim 5)}$ of the NN model for the periodic hill case.
        }
        \label{fig:phill-pfi}
\end{figure}

The PFI values for each input feature of the learned neural network model are illustrated in Figure~\ref{fig:phill-pfi}.
It can be seen that the two invariants $\theta_1$ and $\theta_2$ have large PFI values, and the feature $\nu_\text{t}/\nu$ has relatively small PFI value.
The features $\tilde{\mathcal{P}}/(\beta^*\omega k)$ and $\|\bm{\tau}\|/k$ have almost negligible PFI values.
This suggests using only tabular data of the three important features $\theta_1,\theta_2,\nu_\text{t}/\nu$ for symbolic regression. 
The AI Feynman method gives the tensor coefficient functions as 
\begin{equation}
\label{equ:k-omega-quadratic}
\begin{aligned}
    g^{(1)}(\hat{q}_1,\hat{q}_2,\hat{q}_3)&=-0.108+0.010\bra{\hat{q}_1+\hat{q}_2}\bra{\hat{q}_3-\hat{q}_1-\hat{q}_2},\\
    g^{(2)}&=\num{2.043e-4}g^{(1)}+\num{1.793e-05}.
\end{aligned}
\end{equation}
This symbolic expression differs from the linear eddy viscosity model in the value of $g^{(1)}$. 
Based on the Bradshaw assumption~\cite{Bradshaw1971} that the ratio of Reynolds shear stress to kinetic energy is constant in the log-law region, the linear eddy viscosity model derives $g^{(1)}=-0.09$.
In contrast, the learned model identifies a nonuniform distribution of the ratio in the separated flow over periodic hills.
Moreover, in a uniform shear flow, the mean velocity gradient has only one component, $\partial_y U$, leading to $\hat{q}_1+\hat{q}_2=0$.
Under these conditions, the learned symbolic model can be simplified to $g^{(1)}=-0.108$, which is close to the value used in the linear eddy viscosity model.
This shows the learned symbolic model can effectively degrade to the linear eddy viscosity model in the uniform shear flow.

The PFI method can improve the robustness of the learned model and the efficiency of the symbolic regression.
Specifically, the symbolic method without the PFI can give a complicated model form involving all the six input features, as detailed in~\ref{sec:all-symbol}. 
Such complex models can lead to the divergence of the posterior prediction when coupled with the RANS solver.
 In contrast, the symbolic method with the PFI provides simple model formulations with only the three important features. 
On the other hand, the PFI analysis can exclude the insignificant features prior to the symbolic regression, which can simplify the problem and improve the training efficiency as shown in Table~\ref{tab:addlabel}. 

\begin{figure}[!htb]
        \centering
        \includegraphics[width=0.85\linewidth]{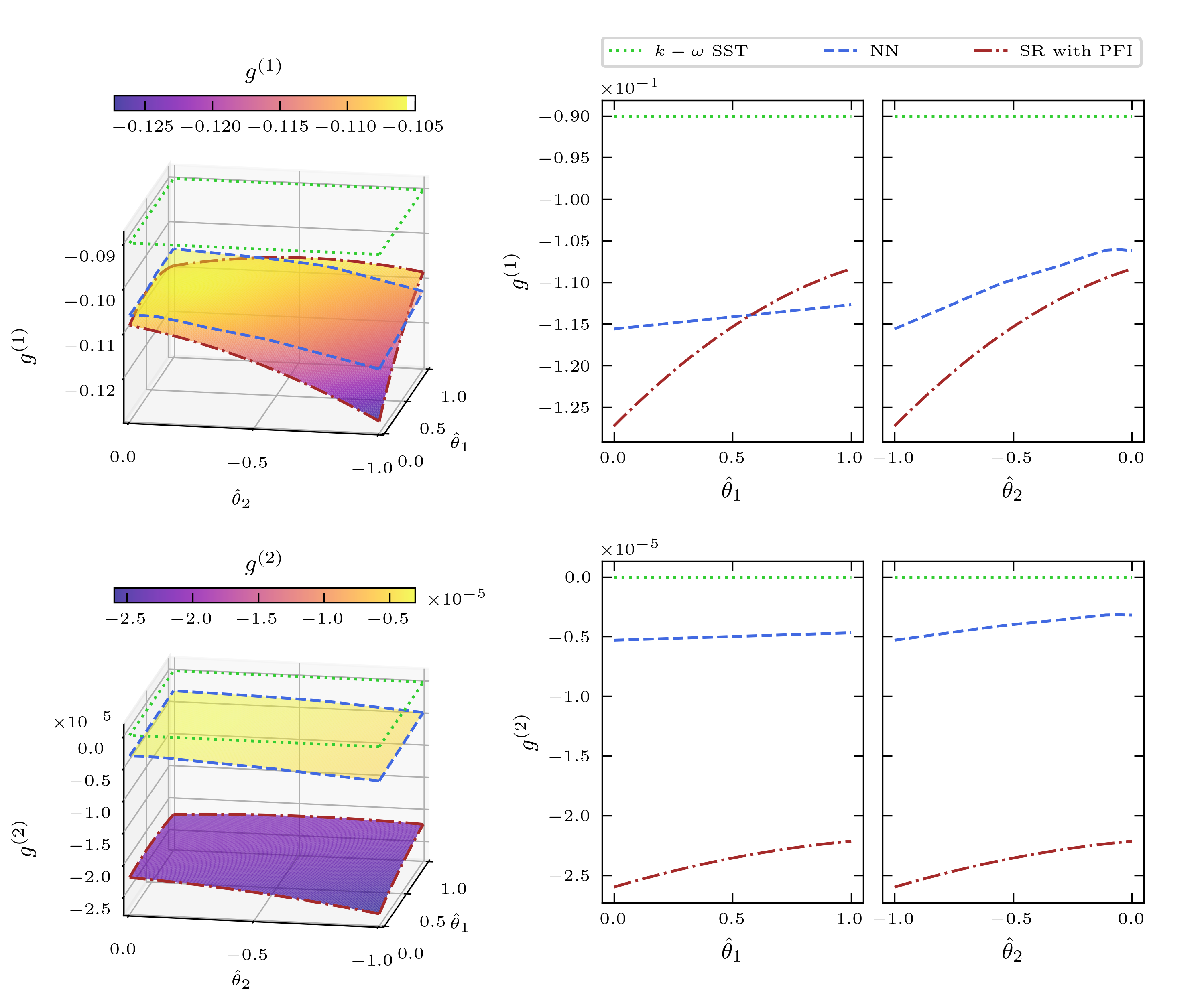}
        \caption{
        (Surface plots) The functional mapping from the scalar invariants $\hat{\theta}_1,\hat{\theta}_2$ to the tensor coefficients $g^{(1)},g^{(2)}$ at the mean value of $\hat{q}_3$ with comparison among the baseline model---$k$-$\omega$ SST (green dotted edge), the neural network model (blue dashed edge), and the SR with PFI (red dotdashed edge) for the periodic hill case. (Curve plots) Plots of the functional mapping at slices of $\hat{\theta}_2=-1$ and $\theta_1=0$, respectively.}
        \label{fig:map}
\end{figure}

The mapping from $\hat{q}_1,\hat{q}_2$ to the tensor coefficients $g^{(1)},g^{(2)}$ is illustrated on the left of Figure~\ref{fig:map}, where $\hat{q}_3$ is fixed at the mean value across the field.
The slices at $\hat{q}_2=-1$ and $\hat{q}_1=0$ are also presented on the right of Figure~\ref{fig:map}.
As depicted in the plots, both the learned neural network and the symbolic model increase the magnitude of the tensor coefficients compared to the baseline model.
Also, the functions $g^{(1)}$ and $g^{(2)}$ exhibit only minor variations for the range of large $\hat{q}_1$ and small $\hat{q}_2$, which is consistent with the neural network model learned by the previous study~\cite{zhang2022}.

\subsubsection{Generalization test}

\begin{figure}[!htb]
            \centering
            \subfigure[Periodic hills of different slopes.]{\label{subfig-pehills}\includegraphics[]{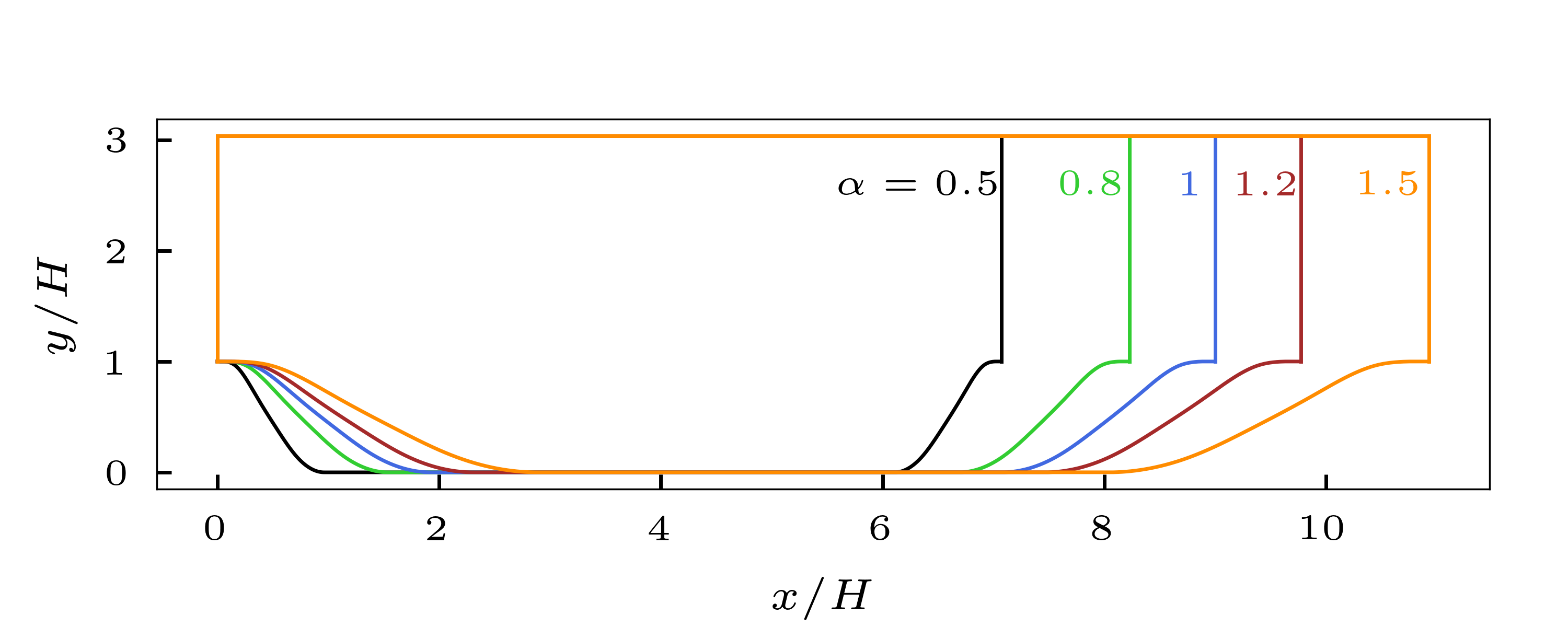}}
            \subfigure[Rounded backward-facing step.]{\label{subfig-rbfs}\includegraphics[]{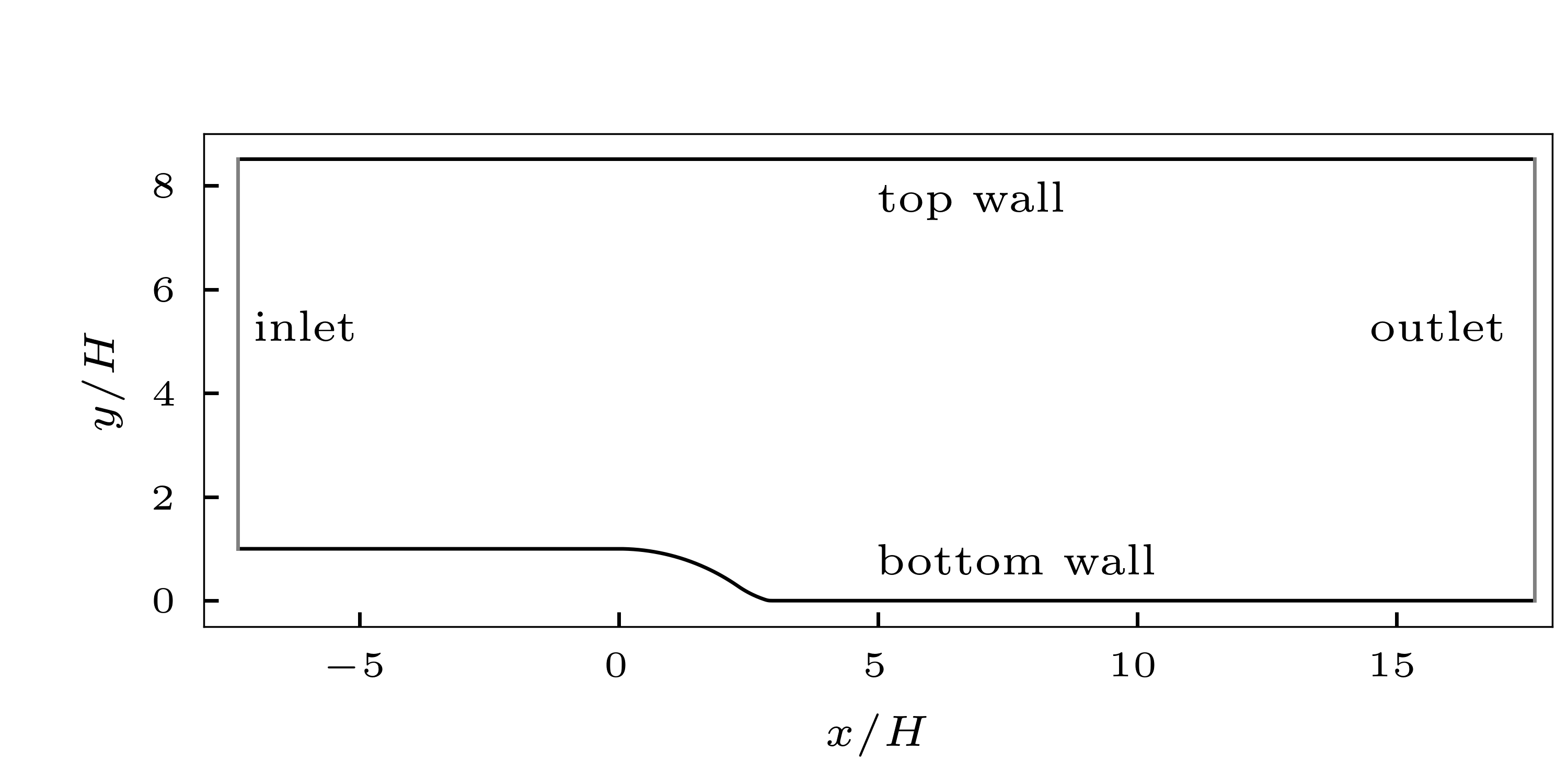}}
            \caption{
            Configurations for the generalization test: (a) the geometry of the five periodic hill cases, with varying slopes, detailed in \cite{XIAO2020104431}, with a Reynolds number based on hill height and bulk velocity of $Re_H=5600$; 
            (b) the geometry and mesh for the curved backward-facing step, with a Reynolds number based on the mean inlet velocity and step height of $Re_H=13700$.}
            \label{fig:valid-cases}
    \end{figure}

In this section, we examine the generalization capability of the learned symbolic models for similar geometries to the training case.
The test cases include periodic hills with different slopes and a curved back-forward step.
The geometries of these cases are illustrated in Figure~\ref{fig:valid-cases}.

The DNS results~\cite{XIAO2020104431} are available for parameterized periodic hills of different slopes.
 The slope is determined by the steepness ratio $\alpha$, which quantifies the relationship between the total horizontal length $L_x$ and the height of the hills~$H$ as $L_x/H=3.858\alpha+5.142$.
The height of the hills remains constant, resulting in varying lengths for different slopes.
Figure~\ref{subfig-pehills} illustrates the schematic of the periodic hill for slopes of $\alpha=0.5,0.8,1.0,1.2,1.5$.
The case with $\alpha=1$ is used for model training.
 The Reynolds number on bulk velocity and crest height is $Re_H=5600$.

\begin{figure}[!htb]
    \centering
    \subfigure[Slope parameter $\alpha=0.5$.]{\includegraphics[]{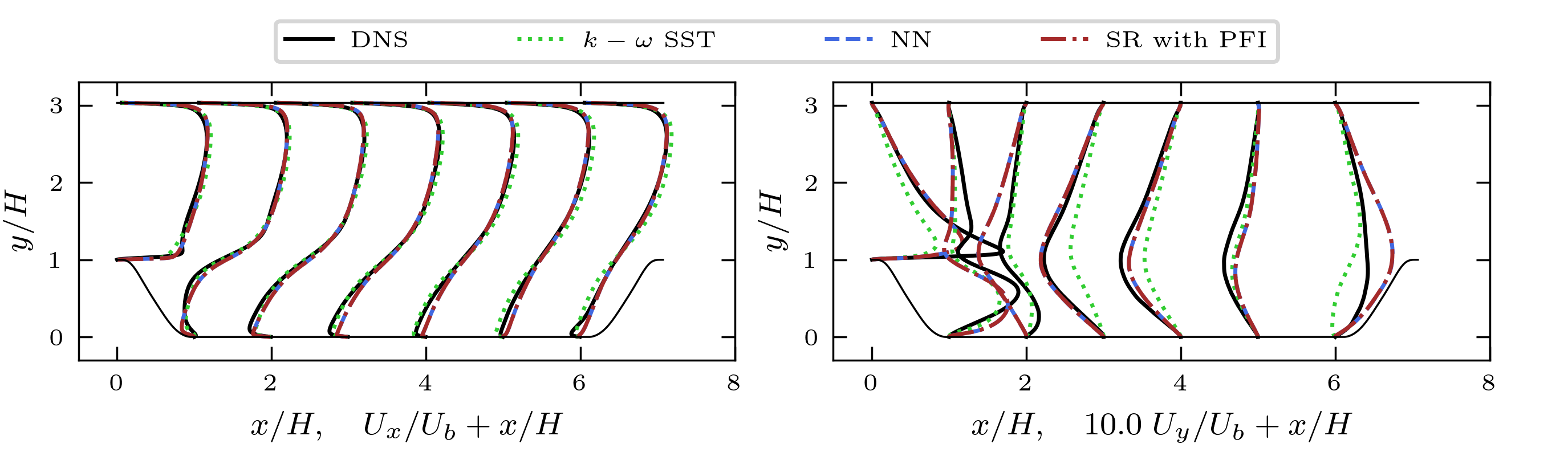}}
    \subfigure[Slope parameter $\alpha=0.8$.]{\includegraphics[]{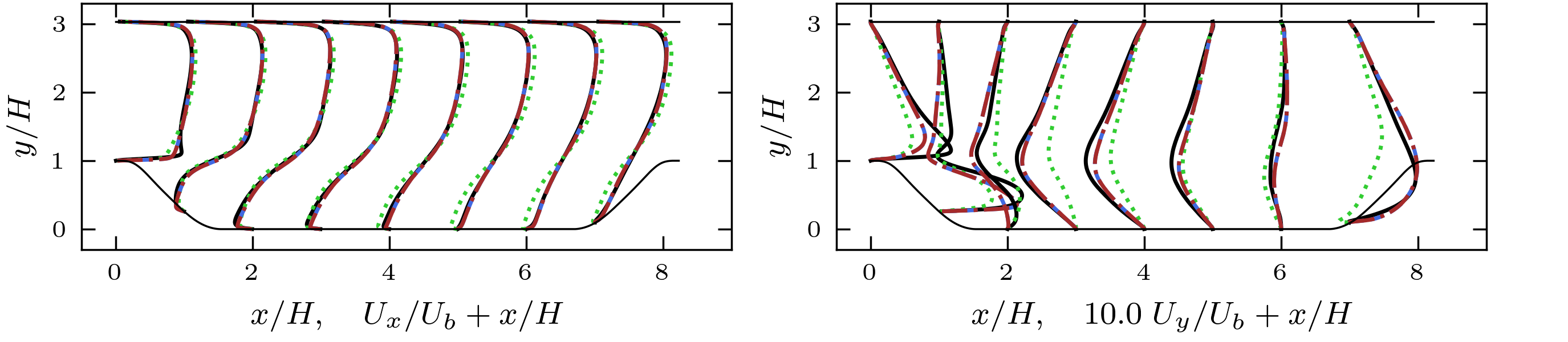}}
    \subfigure[Slope parameter $\alpha=1.2$.]{\includegraphics[]{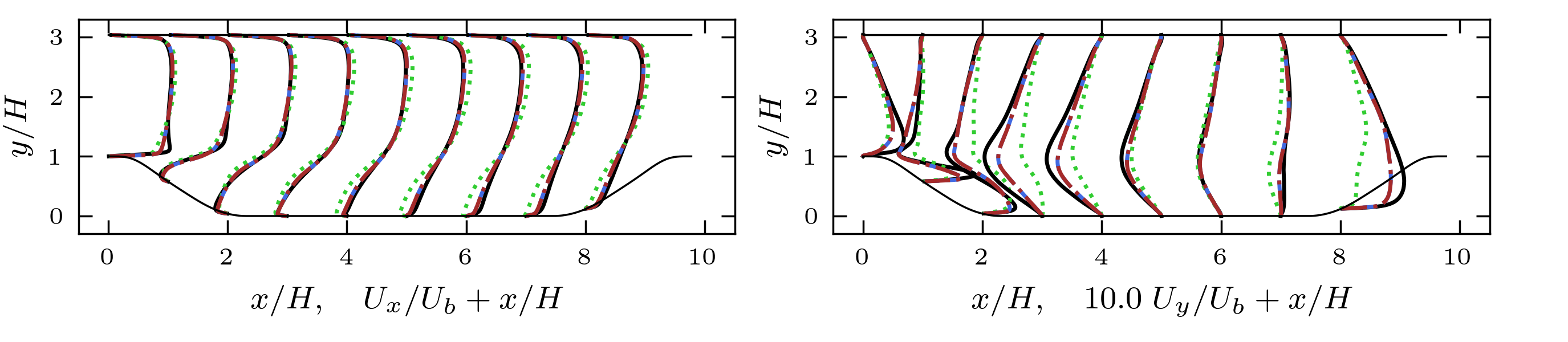}}
    \subfigure[Slope parameter $\alpha=1.5$.]{\includegraphics[]{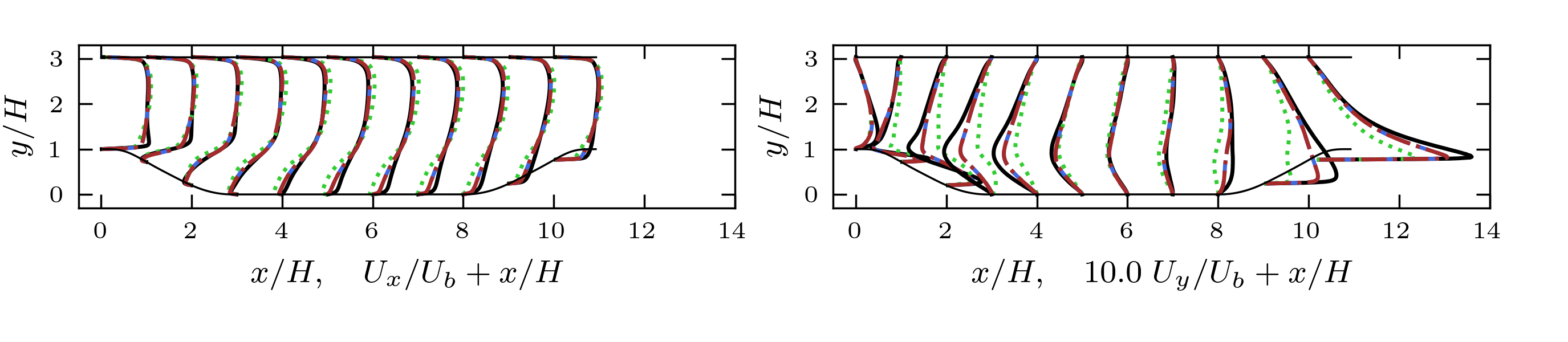}}
    \caption{Comparison of mean velocity in the stream-wise and wall-normal directions along profiles predicted by the baseline model ($k$-$\omega$ SST), the neural network (NN) model, the symbolic regression (SR) model with PFI, and the DNS results for flows over periodic hills over different slopes. 
    }
    \label{fig:phills-slopes-velocity}
\end{figure}

 Figure \ref{fig:phills-slopes-velocity} compares the mean velocity profiles of periodic hills among the baseline model, the DNS results, the neural network model, and the symbolic model. 
 The baseline model consistently overestimates streamwise velocity near the upper wall and underestimates in the valley near the bottom wall.
 For the vertical velocity~$U_y$, the model underestimates the velocity magnitude in most areas.
 In contrast, the symbolic model provides better predictions that align closely with the DNS data across all slopes by increasing the streamwise velocity near the bottom wall and the vertical velocity.
 Moreover, the learned symbolic model makes consistent predictions with the neural network model, although the learned model functions have noticeable differences as presented in Figure~\ref{fig:map}.

We also test the learned symbolic model in flow over the curved backward-facing step.
The geometry is shown in Figure \ref{subfig-rbfs}.
The range of the computational domain is $22.7H \times 9.48H$, where $H$ is the height of the rounded step. 
The inlet is located at $x/H=-7.34$, and the step begins at $x/H=0$ with a streamwise length of $2.937H$~\cite{Yacine2012}. 
 The Reynolds number based on the inlet bulk velocity and step height is $Re_H=13700$.
 No-slip boundary conditions are imposed at the top and bottom walls.
The quadrilateral mesh size is about \num{37000} and is refined near walls.
 A fully developed velocity profile is prescribed at the inlet, and a zero-gradient boundary condition is applied at the outlet.
 Also, the zero pressure gradient is applied at the inlet, and the Dirichlet boundary condition is imposed at the outlet.
A two-dimensional mesh is utilized for the RANS simulation, considering the flow to be statistically homogeneous in the spanwise direction.
 The large-eddy simulation (LES) predictions~\cite{Yacine2012} are used to evaluate the accuracy of the learned models.

\begin{figure}[!htb]
    \centering
    \includegraphics{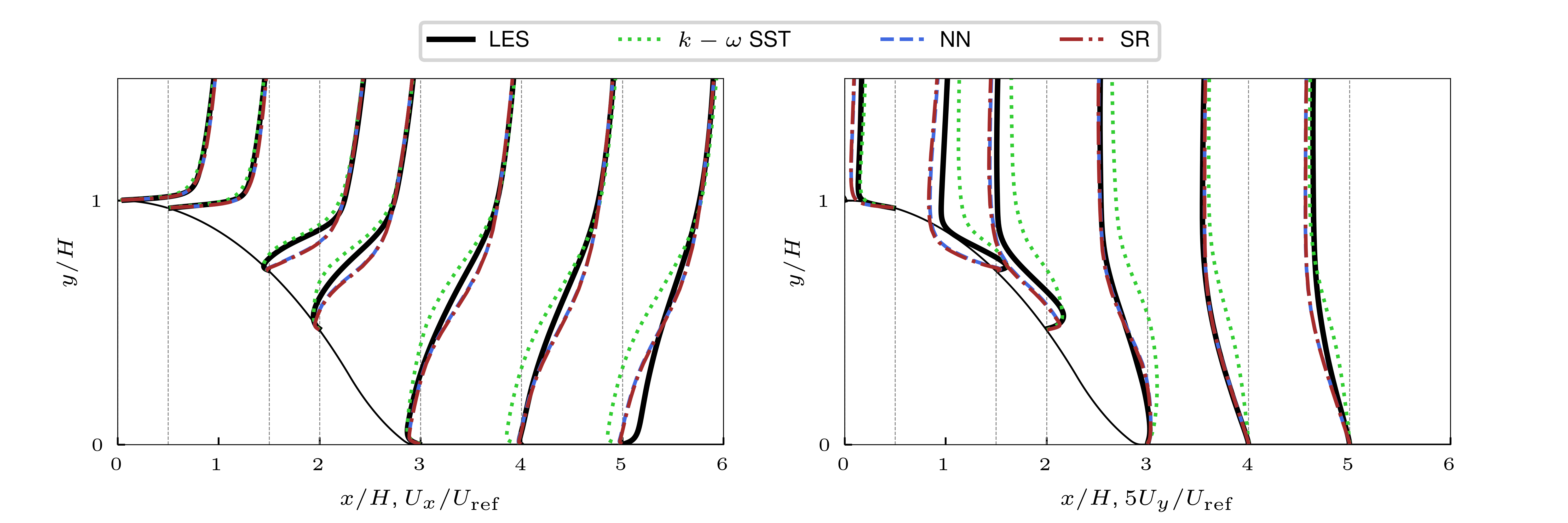}
    \caption{Comparison of mean velocity profiles at $x/H=0,0.5,1.5,2,3,4,5$ among the baseline model ($k$-$\omega$ SST), the LES result~\cite{Yacine2012}, the neural network (NN) model, and the symbolic regression (SR) model for flows over of a curved backward-facing step.}
    \label{fig:rbfs-mean-velocity}
\end{figure}

 The stream-wise and wall-normal velocity profiles at $x/H=0,0.5,1.5,2,3,4,5$ of the learned models are presented in Figure \ref{fig:rbfs-mean-velocity} with comparison to the LES results.
 These locations are selected to capture important flow features, i.e., the attached boundary layer at $x/H = 0$ and $0.5$, the flow recirculation at $x/H = 1.5, 2, 3,$ and $4$, and the post-reattachment flow at $x/H = 5$.
 The velocity profiles exhibit typical flow separation but with notable differences among the different models. 
 Specifically, the reverse-flow region is thin based on the LES results~\cite{Yacine2012}, with a maximum thickness of $0.33H$ at $x/H=2.9$ and a maximum velocity of $13\%$ at $x/H=3$.
 The baseline $k$-$\omega$ SST model predicts a thicker reverse-flow region extended downstream at $x/H=4$ and $5$, where there is no reverse flows based on the LES prediction.
 In contrast, the symbolic model predicts no reverse flow in this region, which is consistent with the LES results. 
 As for the vertical velocity~$U_y$, the learned symbolic model significantly improves the prediction at $x/H=3, 4,$ and $5$.
 The result shows the superiority of the symbolic model in predicting the reverse flow over the curved back-forward step, compared to the $k$-$\omega$ SST model.

To thoroughly investigate the consistency of posterior performance for both the neural network and symbolic models, we train eight neural networks, each initialized with different samples of weights.
Specifically, we generate eight groups of samples for learning neural network models.
Each group is sampled using distinct random seeds from a Gaussian distribution with initial neural network weights as the mean.
The sampling process is detailed in \ref{sec:procedure}. 
The sample mean of the neural network inputs and outputs from these training sessions is then used to identify a symbolic model. 
The symbolic model derived from the averaged neural network inputs and outputs shares the same functional form as Eq.\eqref{equ:k-omega-quadratic}, but with slightly different parameters: $g^{(1)}=-0.110+0.011(\hat{q}_1+\hat{q}_2)(\hat{q}_3-\hat{q}_1-\hat{q}_2),g^{(2)}=\num{1.972e-4}g^{(1)}+\num{1.773e-5}$. 
The mean and standard deviation (std.) of the prediction errors for the neural network models are provided in Table~\ref{tab:pehills-misfit}, alongside the errors of the baseline $k$-$\omega$ SST model and the symbolic expression model.
In all cases, both the neural network model and the symbolic model provide improved predictions compared to the baseline model. 
Specifically, for periodic hills with $\alpha=0.8$, $1.2$, and $1.5$, the baseline model leads to large discrepancies of $11.866\%$, $17.544\%$, and $21.14\%$, respectively, while both the neural network and symbolic model can reduce the predictive errors below $10\%$.
For the periodic hill of $\alpha=0.5$ and the curved backward-facing step, the learned models marginally improve the velocity predictions. 
The posterior tests show that the learned symbolic model can be well generalized to flows in similar geometries to the training case.
For periodic hill cases with slopes of $\alpha=0.5$, $0.8$, and $1.0$, the symbolic model underperforms compared to the neural network model.
However, for periodic hills with flatter slopes and the curved backward-facing step case, the symbolic model can have better predictive performance than the neural network model.

\begin{table}[!htb]
    \centering
    \caption{Prediction errors of the baseline model, the neural network (NN) model, and the symbolic regression (SR) model with the PFI analysis.}
    \resizebox{\linewidth}{!}{
    \begin{tabular}{lcccc}
    \toprule
          case  & $k$-$\omega$ SST (\%) & mean NN (\%) & std. NN (\%) & SR with PFI (\%)  \\ 
    \midrule
          periodic hills $\alpha=0.5$ & $\num{8.507}$ & \num{7.865}& \num{0.0905} & \num{8.493} \\
          periodic hills $\alpha=0.8$ & $\num{11.866}$ & \num{4.929} & \num{0.0709} & \num{5.432} \\
          periodic hills $\alpha=1.0$ (training case) & $\num{13.556}$  & \num{4.312}& \num{0.0761} & \num{4.629} \\
          periodic hills $\alpha=1.2$ & $\num{17.544}$ & \num{6.128} &\num{0.1377} & \num{5.309} \\
          periodic hills $\alpha=1.5$ & $\num{21.140}$ & \num{8.829} & \num{0.2128} & \num{7.727} \\
          curved backward-facing step & \num{8.803} & \num{7.321} & \num{0.0306} & \num{7.463} \\
    \bottomrule
    \end{tabular}
    }
    \label{tab:pehills-misfit}
\end{table}

\section{Conclusion}\label{sec:IV}

In this work, the symbolic turbulence model is learned from indirect observation data based on the neural network and feature importance analysis.
Specifically, a neural network model is first trained with the indirect velocity observation, which avoids the difficulty in obtaining high-fidelity Reynolds stress data and ensures the consistency between training and prediction environments.
Also, the permutation feature importance (PFI) method is used to identify important input features of the neural network.
These features are subsequently utilized as inputs for physics-inspired symbolic regression, which searches for mathematical formulations of Reynolds stress that can approximate the neural network-represented functional mapping.
The proposed strategy is shown to efficiently learn symbolic turbulence models from indirect observation data via neural network training and feature reduction.

The feasibility of the proposed method is first validated using flows in a square duct, where the learned symbolic model has similar expression forms to the true analytical model.
Further, the method is tested for flows over periodic hills with mean velocity profiles based on velocity data from DNS at limited locations.
The results show that the learned symbolic model can consistently improve the velocity predictions compared to the baseline $k$-$\omega$ SST model for similar flow configurations, including periodic hills with different steepness ratios and the curved backward-facing step.
Both cases highlight the enhanced efficiency of symbolic turbulence modeling with neural networks and feature importance analysis.

\appendix

\section{The baseline $k$-$\omega$ SST model}
\label{sec:sst}
In the $k$-$\omega$ shear stress transport (SST) model, the transport equations for $k$ and $\omega$ are formulated as
\begin{subequations}
    \begin{equation}
        \frac{\p k}{\p t}+\bm{U}\cdot\nabla k=\tilde{\mathcal{P}}-\beta^* k\omega+\nabla\cdot\sbra{\bra{\nu+\nu_\turb \sigma_k}\nabla k},
    \end{equation}
    \begin{equation}
        \frac{\p\omega}{\p t}+\bm{U}\cdot\nabla\omega=\alpha\mathcal{S}^2-\beta\omega^2+\nabla\cdot\sbra{\bra{\nu+\nu_\turb\sigma_\omega}\nabla\omega}+2\bra{1-F_1}\sigma_{\omega2}\frac{1}{\omega}\nabla k\cdot\nabla\omega,
    \end{equation}
\end{subequations}
where $\nu_\text{t}$ is the turbulent eddy viscosity constructed by
\begin{equation*}
    \nu_\text{t}=\frac{a_1 k}{\max\bra{a_1\omega,\mathcal{S}F_2}} \text{.}
\end{equation*}
 The turbulent kinetic energy production $\tilde{\mathcal{P}}$ is 
\begin{equation*}
    \tilde{\mathcal{P}}=\min\bra{\bm{\tau}:\nabla\bm{U},10\beta^* k\omega},
\end{equation*}
where $\mathcal{S}=\sqrt{S_{ij}S_{ij}}$ is the mean strain rate.
The symbols $F_1,F_2$ represent the blending functions as
\begin{equation*}
\begin{aligned}
    F_1&=\tanh\cbra{\cbra{\min\bra{\max\bra{\frac{\sqrt{k}}{\beta^* \omega y},\frac{500\nu}{y^2\omega}},\frac{4\sigma_{\omega 2}k}{CD_{k\omega} y^2}}}^4},\\
    F_2&=\tanh\cbra{\cbra{\max\bra{\frac{2\sqrt{k}}{\beta^*\omega y},\frac{500\nu}{y^2\omega}}}^2},\\
    CD_{k\omega}&=\max\bra{2\sigma_{\omega 2}\frac{1}{\omega}\nabla k\cdot\nabla \omega,\num{1.e-10}},
\end{aligned}
\end{equation*}
where $y$ is the distance to the nearest wall.
The parameters~$\phi$ are calculated by blending the counterpart of $k$-$\omega$ model ($\phi_1$) and $k$-$\varepsilon$ model ($\phi_2$) as $\phi=\phi_1 F_1+\phi_2 F_2.$ 
The coefficients $\alpha, \beta, \sigma_k, \sigma_\omega$ are blends of $(\alpha_1=5/9, \beta_1=3/40, \sigma_{k1}=0.85, \sigma_{\omega 1}=0.5),(\alpha_2=0.44, \beta_2=0.0828, \sigma_{k2}=1, \sigma_{\omega 2}=0.856)$, respectively. 
Other parameters are set as $\beta^*=0.09, a_1=0.31$.

\section{Procedure of ensemble Kalman method for learning neural network-based turbulence model}
\label{sec:procedure}
Given the initial neural network weight~$\m{w}_0$ and the standard deviations of initial samples and the observation data~$\m{y}$, the ensemble-based neural network training consists of the following steps.
 \begin{itemize}
     \item [(a)] Initial sampling. The neural network weights for each sample are determined by sampling from a Gaussian distribution with a mean of $\m{w}_0$ and a user-specified standard deviation. The initial weight $\m{w}_0$ is pre-trained to make the neural network function exactly as the baseline model by ensuring $g^{(1)}=-0.09$ and the coefficients of other bases equal to zero. 
     The number of samples and the standard deviation value are determined based on our sensitivity study to balance training accuracy and efficiency.
     \item [(b)] Feature extraction. The scalar invariants $\theta_i$ and tensor bases $\m{T}^{(i)}$ are obtained from velocity field $\bm{U}$ and turbulence time scale $1/(C_\mu \omega)$ as Eqs.\eqref{equ:invariants} and \eqref{equ:bases}. 
     The scalar invariants are then locally normalized  to have a range in $[-1,1]$ as inputs for the neural network.
     \item [(c)] Model evaluation. At each time step of the RANS simulation, the Reynolds stress is constructed by combining the bases $\m{T}^{(i)}$ and coefficients $g^{(i)}$ that are predicted by a neural network.
     \item [(d)] Forward propagation. With Reynolds stress provided, the velocity $\bm{U}$, turbulent kinetic energy $k$, and turbulence frequency $\omega$ are updated for the next time step by solving RANS equations and the turbulence transport equations.
     \item [(e)] Kalman update. The weights of neural networks are updated with the ensemble Kalman method by analyzing the flow predictions and the experimental measurements, i.e.
     \begin{equation}
         \m{w}_j^{\ell+1}=\m{w}_j^\ell+\m{K}\bra{\m{y}_j-\mathcal{H}\sbra{\m{w}_j^\ell}},
     \end{equation}
     where $\m{w}_j^\ell$ is the weight of neural network for sample $j$ at training iteration step $\ell$, $\m{K}$ is the Kalman gain matrix, $\m{y}$ is the observation data, and $\mathcal{H}\sbra{\m{w}_j^\ell}$ is the model prediction. 
\end{itemize}

\section{Learned symbolic expressions for the periodic hill case}\label{sec:all-symbol}

AI Feynman generates a Pareto frontier, which can provide a spectrum of symbolic formulas ranging from simple to complex, each offering progressively improved accuracy rather than delivering a single, fixed solution~\cite{Silviu2020}. 
This feature allows users to navigate the trade-off between model complexity and predictive precision. 
In our implementation, we choose the formula that achieves the lowest posterior fitting error and incorporates all three important features simultaneously.
In essence, the Pareto frontier of the AI Feynman, combined with our selection criteria, enables the discovery of models that are not only accurate but also interpretable.

The symbolic regression results obtained by AI Feynman with and without PFI assistance for the periodic hill case are presented in Table~\ref{tab:all-equ-pehills_wo}, respectively.
The prior error in Table~\ref{tab:all-equ-pehills_wo} refers to the relative error in the tensor coefficient $g^{(1)}$ between the predictions of the symbolic model and the neural network model. 
This error measures how closely the symbolic expression approximates the neural network. 
The posterior error refers to the relative error in the mean velocity at observation positions between the predictions of the symbolic turbulence model and the DNS. 
This error assesses the ability of the turbulence model to predict the mean velocity.
 The results in Table~\ref{tab:all-equ-pehills_wo} show that the symbolic models learned without the PFI analysis lead to divergent predictions.
 In contrast, the symbolic regression with the PFI analysis can provide a simple formula of $g^{(1)} = 0.01\left(\hat{q}_1+\hat{q}_2\right) \left(-\hat{q}_1-\hat{q}_2+\hat{q}_3\right)-0.108$ which meet the criteria and have good posterior prediction accuracy.
 Note that the symbolic regression method can also offer expressions with partial features that could make improved posterior predictions but at a relatively high computational cost.
 Besides, it may exclude key flow features in the model formulation, which would deteriorate the generalizability of the learned turbulence model.

\section{Dependence on noise level of training data}\label{sec:noise-check}

\begin{figure}[!htb]
    \centering
    \includegraphics[]{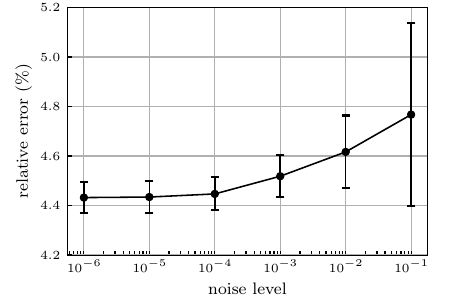}
    \caption{The relative prediction error of the neural network (NN) model with different relative standard deviations of training data.}
    \label{fig:noise-check}
\end{figure}

The ensemble Kalman method inherently takes random noise in the training data into account~\cite{zhang2020evaluation}, which can effectively avoid data overfitting~\cite{zhang2022}. 
To investigate the robustness of the ensemble Kalman method, we test the effect of the data noise on the training performance in the periodic hill case with a steepness ratio of $\alpha=1$.
We vary levels of data noise, i.e., relative standard deviations ranging from $\num{1.e-6}$ to $\num{1.e-1}$, equally spaced on log space with interval $\num{1.e1}$. 
The relative error of mean velocity predicted by the trained neural network models is shown in~Figure \ref{fig:noise-check}. 
The results show that the trained models remain robust to noise levels up to $\num{1.e-2}$, with relatively small prediction errors and sample deviations.
However, as the noise level increases to $\num{1.e-1}$, both the prediction error and the standard deviation increase significantly. 

\begin{table}[!htb]
    \centering
    \begin{tabular}{llll}
    \toprule
    Function $100 g^{(1)}$ &  priori error  & posterior error \\
    \midrule
    $f_1=\left(\hat{q}_1+\hat{q}_2\right) \left(-\hat{q}_1-\hat{q}_2+\hat{q}_3\right)-10.839$ & \num{1.593e-02 } & \num{4.407e-02 } \\
    $f_2=\hat{q}_1+\frac{\hat{q}_2}{-\hat{q}_3+\pi -1}-10.950$ & \num{1.605e-02} & \num{4.456e-02 } \\
    $f_3=-\frac{\left| \hat{q}_1+\hat{q}_2\right|}{\pi}+\hat{q}_1+\hat{q}_2-10.829$ & \num{ 1.622e-02  } & \num{4.411e-02} \\
    $f_4=\hat{q}_1+\frac{-\hat{q}_2+\hat{q}_3-\cos \hat{q}_2}{\hat{q}_1-3}-11.134$ & \num{1.633e-02 } & \num{4.458e-02 } \\
    $f_5=\hat{q}_2+\hat{q}_1 \left(-\hat{q}_3+\pi -1\right)-10.971$   & \num{1.689e-02 } & \num{4.321e-02 } \\
    $f_6=\frac{1}{\hat{q}_3-e^{\hat{q}_1+\hat{q}_2+1}}-10.260$   & \num{1.779e-02 } & \num{4.426e-02} \\
    {\tiny $h_1=\log \left(\frac{\num{1.672e-5}}{\left|1+ \log \hat{q}_3\right| }\right) \left(\tan^{-1}\left(0.368 +\frac{1}{\frac{\frac{\hat{q}_2-\cos \hat{q}_4}{\log (\hat{q}_4+1)+1}-1}{\sqrt{\hat{q}_1+1}}-1}\right)+1\right)$}  & \num{1.591e-02} & \texttt{NAN} \\
    {\scriptsize $h_2=\log \left(\frac{\num{1.672e-5}}{\left| 1+\log \hat{q}_3\right| }\right) \left(\frac{1}{\frac{\frac{\hat{q}_2-\cos \hat{q}_4}{\log (\hat{q}_4+1)+1}-1}{\sqrt{\hat{q}_1+1}}-1}+0.332 \left(\pi -\frac{1}{\hat{q}_5+2 \pi }\right)+0.368\right)$}  & \num{1.583e-02} & \texttt{NAN}\\
    {\scriptsize $h_3=\log \left(\frac{\num{1.672e-5}}{\left| 1+ \log \hat{q}_3\right| }\right) \left(\frac{1}{\frac{\frac{\hat{q}_2-\cos \hat{q}_4}{\log (\hat{q}_4+1)+1}-1}{\sqrt{\hat{q}_1+1}}-1}+0.326 \left(\pi -\frac{1}{\pi  (\hat{q}_5+\pi )}\right)+0.368\right)$}  & \num{1.583e-02} & \texttt{NAN} \\
    {\scriptsize $h_4=\log \left(\frac{\num{1.672e-5}}{\left| 1+\log \hat{q}_3\right| }\right) \left(\sin \left(0.368 +\frac{1}{\frac{\frac{\hat{q}_2-\cos \hat{q}_4}{\log (\hat{q}_4+1)+1}-1}{\sqrt{\hat{q}_1+1}}-1}\right)+0.342 \left(\pi -\frac{1}{\sqrt{\hat{q}_5+1}+\pi }\right)\right)$} & \num{1.583e-02} & \texttt{NAN} \\
    {\scriptsize $h_5=\log \left(\frac{\num{1.672e-5}}{\left| 1+\log \hat{q}_3\right| }\right) \left(\tan ^{-1}\left(0.368 +\frac{1}{\frac{\frac{\hat{q}_2-\cos \hat{q}_4}{\log (\hat{q}_4+1)+1}-1}{\sqrt{\hat{q}_1+1}}-1}\right)+0.342 \left(\pi -\frac{1}{\sqrt{\hat{q}_5+1}+\pi }\right)\right)$}  & \num{1.583e-02} & \texttt{NAN} \\
    \bottomrule
    \end{tabular}
    \caption{The symbolic expressions obtained by AI Feynman for the periodic hill case include functions $f_i,(i=1,2,3,4,5,6)$ with PFI assistance and functions $h_i,(i=1,2,3,4,5)$ without PFI assistance.}
    \label{tab:all-equ-pehills_wo}
\end{table}

\section*{Acknowledgements}
This work is supported by the NSFC Basic Science Center Program for `Multiscale Problems in Nonlinear Mechanics' (No. 11988102), the National Natural Science Foundation of China (No. 12102435), CAS Project for Young Scientists in Basic Research(YSBR-087) and the Young Elite Scientists Sponsorship Program by CAST (No. 2022QNRC001). 
The authors also would like to thank the reviewers for their constructive and valuable comments, which helped improve the quality and clarity of this manuscript.

\bibliography{refs}

\end{document}